\documentclass[twocolumn,aps,pra,superscriptaddress,showpacs,tightenlines]{revtex4-1}
\usepackage{amsmath}
\usepackage{amsfonts}
\usepackage{graphicx}
\usepackage{units}
\usepackage{color}
\usepackage[colorlinks,linkcolor=blue,anchorcolor=blue,citecolor=blue,urlcolor=blue]{hyperref}
\usepackage{multirow}

\begin{document}
\title{Multiphoton blockade in the two-photon Jaynes-Cummings model}

\author{Fen Zou}
\affiliation{Key Laboratory of Low-Dimensional Quantum Structures and Quantum Control of
Ministry of Education, Key Laboratory for Matter Microstructure and Function of Hunan Province, Department of Physics and Synergetic Innovation Center for Quantum Effects and Applications, Hunan Normal University, Changsha 410081, China}

\author{Xiao-Ya Zhang}
\affiliation{Key Laboratory of Low-Dimensional Quantum Structures and Quantum Control of
Ministry of Education, Key Laboratory for Matter Microstructure and Function of Hunan Province, Department of Physics and Synergetic Innovation Center for Quantum Effects and Applications, Hunan Normal University, Changsha 410081, China}

\author{Xun-Wei Xu}
\affiliation{Key Laboratory of Low-Dimensional Quantum Structures and Quantum Control of
Ministry of Education, Key Laboratory for Matter Microstructure and Function of Hunan Province, Department of Physics and Synergetic Innovation Center for Quantum Effects and Applications, Hunan Normal University, Changsha 410081, China}
\affiliation{Department of Applied Physics, East China Jiaotong University, Nanchang 330013, China}

\author{Jin-Feng Huang}
\email{jfhuang@hunnu.edu.cn}
\affiliation{Key Laboratory of Low-Dimensional Quantum Structures and Quantum Control of
Ministry of Education, Key Laboratory for Matter Microstructure and Function of Hunan Province, Department of Physics and Synergetic Innovation Center for Quantum Effects and Applications, Hunan Normal University, Changsha 410081, China}

\author{Jie-Qiao Liao}
\email{jqliao@hunnu.edu.cn}
\affiliation{Key Laboratory of Low-Dimensional Quantum Structures and Quantum Control of
Ministry of Education, Key Laboratory for Matter Microstructure and Function of Hunan Province, Department of Physics and Synergetic Innovation Center for Quantum Effects and Applications, Hunan Normal University, Changsha 410081, China}

\date{\today}

\begin{abstract}
We study multiphoton blockade and photon-induced tunneling effects in the two-photon Jaynes-Cummings model, where a single-mode cavity field and a two-level atom are coupled via a two-photon interaction. We consider both the cavity-field-driving and atom-driving cases, and find that single-photon blockade and photon-induced tunneling effects can be observed when the cavity mode is driven, while the two-photon blockade effect appears when the atom is driven. For the atom-driving case (the two-photon transition process), we present a criterion of the correlation functions for the multiphoton blockade effect. Specifically, we show that quantum interference can enhance the photon blockade effect in the single-photon cavity-field-driving case. Our results are confirmed by analytically and numerically calculating the correlation function of the cavity-field mode. Our work has potential applications in quantum information processing and paves the way for the study of multiphoton quantum coherent devices.
\end{abstract}
\maketitle

\section{Introduction \label{intro}}

The photon blockade (PB) effect~\cite{imamoglu1997Strongly}, as a typical photon correlation phenomenon, not only has significance in the study of the fundamentals of quantum optics, but also possesses wide applications in modern quantum devices and quantum information science. So far, there exist two kinds of PB, conventional photon blockade (CPB) and unconventional photon blockade (UPB), which are based on different physical mechanisms. The former is caused by the nonlinearity in the energy spectrum, while the latter is induced by the quantum interference effect between different transition channels. In CPB, the capture of a single photon in a nonlinear system blocks the excitation of the second and subsequent photons. Thus, a sequence of single photons can be generated and such systems can be implemented as single-photon source devices. In this sense, PB can change a classical light field into a nonclassical light field. In general, the signatures of PB can be observed from photon antibunching and sub-Poissonian photon-number statistics.

In recent years, great advances have been made in the topic of PB. On one hand, the CPB effect has been theoretically investigated in a variety of quantum systems, e.g., cavity quantum electrodynamic (QED) systems~\cite{tian1992Quantum,
birnbaum2005Photon,faraon2008Coherent,faraon2010Generation,reinhard2012Strongly,ridolfo2012Photon,peyronel2012Quantum,bajcsy2013Photon,
muller2015Coherent,radulaski2017Photon,wang2017Phasemodulated,pietikainen2017Observation,han2018Electromagnetic,trivedi2019Photona,hou2019Interfering}, circuit-QED systems~\cite{hoffman2011Dispersive,lang2011Observationa,liu2014Blockade}, Kerr-type nonlinear cavities~\cite{leonski1994Possibility,imamoglu1997Strongly,ferretti2010Photon,liao2010Correlateda,ghosh2019Dynamicala}, optomechanical systems~\cite{rabl2011Photon,liao2013Photon,liao2013Correlated,komar2013Singlephoton,wang2015Tunable,zhu2018Controllable,zou2019Enhancement}, and other systems~\cite{zheng2011CavityFree,majumdar2013Singlephoton,huang2013Photon,zeytinoglu2018Interactioninduced,zou2018Photon,pietikainen2019Photon}. The CPB effect has also been experimentally demonstrated with a single atom trapped in an optical cavity~\cite{birnbaum2005Photon}, a quantum dot in a photonic crystal cavity~\cite{faraon2008Coherent,reinhard2012Strongly,muller2015Coherent}, and a single superconducting artificial atom coupled to a microwave transmission-line resonator~\cite{hoffman2011Dispersive,lang2011Observationa}. On the other hand, the UPB effect has been theoretically studied in coupled Kerr-type nonlinear cavities~\cite{liew2010Single,bamba2011Origin,xu2014Tunable,shen2015Tunable,flayac2017Unconventional,wang2018Photon,ryou2018Strong}, cavity-QED systems~\cite{
zhang2014Optimala,tang2015Quantum,liu2016Mode}, coupled optomechanical systems~\cite{xu2013Antibunching,sarma2018Unconventional,li2019Nonreciprocal}, and other systems~\cite{lemonde2014Antibunching,gerace2014Unconventional,zhou2015Unconventional,zhou2016Strong,sarma2017Quantuminterferenceassisted}. The UPB effect has also been experimentally demonstrated in an optical microcavity coupled to a single semiconductor quantum dot~\cite{snijders2018Observation} and in a superconducting circuit consisting of two coupled resonators~\cite{vaneph2018Observation}.

Previous studies on PB have been aimed mainly at single-photon blockade (1PB). Most recently, two-photon blockade (2PB) has been experimentally~\cite{hamsen2017TwoPhoton} and theoretically~\cite{shamailov2010Multiphoton,miranowicz2013Twophoton,miranowicz2014Statedependent,
hovsepyan2014Multiphoton,carmichael2015Breakdown,deng2015Enhancement,zhu2017Collective,huang2018Nonreciprocal,bin2018Twophoton,lin2019Manipulation,
villas-boas2019Multiphoton,kowalewska-kudlaszyk2019Twophoton} investigated in various configurations. Two-photon blockade means that the resonance absorption of two photons in a nonlinear system will suppress the transmission of subsequent photons. Such systems with 2PB can be used for two-photon source devices. In addition, photon-induced tunneling (PIT) with photon bunching has also been explored in a photonic crystal cavity coupled to a quantum dot~\cite{faraon2008Coherent,majumdar2012LossEnabled,majumdar2012Probing,rundquist2014Nonclassical}, optomechanical systems~\cite{xu2013Photoninduced}, and  other systems~\cite{huang2018Nonreciprocal,kowalewska-kudlaszyk2019Twophoton}, i.e., the absorption of the first photon favors that of the second or subsequent photons. PIT has been observed experimentally in Refs.~\cite{faraon2008Coherent,majumdar2012LossEnabled,rundquist2014Nonclassical}.
\begin{figure}
\center
\includegraphics[width=0.47 \textwidth]{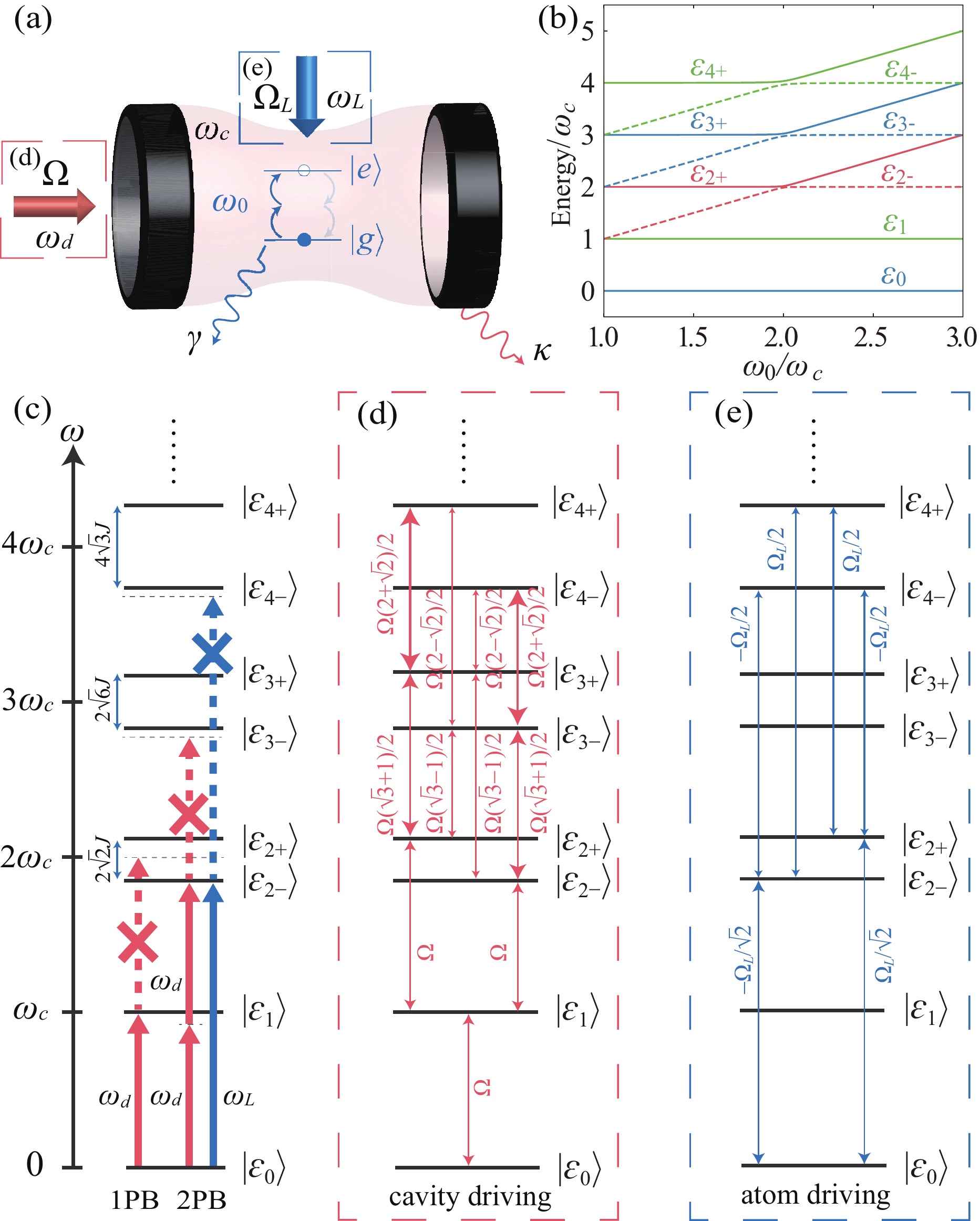}
\caption{(Color online) (a) Schematic of the two-photon JC model, which is composed of a single-mode cavity field coupled to a two-level atom via a two-photon interaction. (b) The energy spectrum of the Hamiltonian $\hat{H}_{\text{2pJC}}$ as a function of the ratio $\omega_{0}/\omega_{c}$ in the subspace associated with zero, one, two, three, and four photons at $J/\omega_{c}=0.01$. (c) Energy-level schematic of the system explaining the occurrences of 1PB and 2PB in the resonant case $\omega_{0}=2\omega_{c}$. The system can be excited by driving either the cavity [(d) the transition processes (red)] or the atom [(e) the transition processes (blue)].}
\label{Fig1}
\end{figure}

Based on the physical picture of multiphoton blockade, a natural question is: what is the influence of the multiphoton physical transition and the multiphoton driving processes on the multiphoton blockade effect? To study this question, in this work we study the multiphoton blockade effect in the two-photon Jaynes-Cummings (JC) model~\cite{sukumar1981Multiphonon,singh1982Field,gerry1988Twophoton,peng1993Influence,ng1999Exact}, which describes the two-photon interaction of a single bosonic mode with a two-level system. This model has been widely studied in quantum optics and quantum information sciences~\cite{travenec2012Solvability,chen2012Exact,dong2012Quantumstate,felicetti2015Spectral,puebla2017Protected,felicetti2018Twophoton}. The strong nonlinearity induced by the interaction gives rise to many important quantum effects at the level of few photons. Here we demonstrate the PB effects in this system by driving either the cavity field or the atom. For this system, the cavity-field driving and the atom driving will induce different photon transition processes. Corresponding to single-photon cavity-field driving and atom driving, two-photon injection processes are, respectively, the second-order single-photon process and the first-order two-photon process. In the single-photon cavity-field-driving case, we find that 1PB and PIT effects can occur in this system, while the 2PB effect cannot appear. In particular, the 1PB effect induced by the destructive quantum interference between the two different paths can also be observed in the off-resonant case. Furthermore, we investigate the photon statistics in the atom-driving case by numerically calculating the correlation function of the cavity field. We find that the 2PB effect can be observed when the atom is driven, while the 1PB effect cannot occur. We also discuss the two-photon cavity-field-driving case and find that the 2PB effect cannot appear.

The rest of this paper is organized as follows. In Sec.~\ref{modelsec}, we introduce the two-photon JC model. In Sec.~\ref{PBPIT}, we present the criteria for the $n$PB and PIT effects. In Secs.~\ref{cavdrsec} and~\ref{atdrsec}, we study photon blockade effects in the cavity-field-driving and atom-driving cases, respectively. The Discussion and Conclusion are given in Secs.~\ref{Discussions} and~\ref{conclusion}, respectively. For completeness, we present an Appendix on the derivation of the two-photon JC Hamiltonian in a superconducting circuit.

\section{Model \label{modelsec}}

We consider a two-photon JC model [Fig.~\ref{Fig1}(a)], which is composed of a single-mode cavity field coupled to a two-level atom via a two-photon physical interaction~\cite{felicetti2018Twophoton}. The Hamiltonian of the two-photon JC model reads ($\hbar=1$)
\begin{equation}
\hat{H}_{\text{2pJC}}=\omega_{c}\hat{a}^{\dagger}\hat{a}+\omega_{0}\hat{\sigma}_{+}\hat{\sigma}_{-}+J(\hat{a}^{\dagger2}\hat{\sigma}_{-}+\hat{\sigma}_{+}\hat{a}^{2}), \label{Ham}
\end{equation}
where $\hat{a}^{\dagger}$ and $\hat{a}$ are, respectively, the creation and annihilation operators of the single-mode cavity field with resonance frequency $\omega_{c}$. The operators $\hat{\sigma}_{+}=\vert e\rangle\langle g\vert$ and $\hat{\sigma}_{-}=\vert g\rangle\langle e\vert$ are, respectively, the raising and lowering operators of the two-level atom with an energy separation $\omega_{0}$ between the excited state $\vert e\rangle$ and the ground state $\vert g\rangle$. The last term in Eq.~(\ref{Ham}) denotes the two-photon interaction between the cavity field and the two-level atom with coupling strength $J$. Note that the two-photon JC model can be implemented with superconducting circuits, in which the two-level system can be either a flux qubit~\cite{felicetti2018Twophoton} or a split-Cooper-pair-box charge qubit~(see Appendix).

It is generally known that the CPB effect is caused by anharmonicity of the eigenenergy spectrum. To study the PB effect in the two-photon JC model, in the following we calculate its eigensystem and analyze its energy spectrum. In the two-photon JC model, the weighted excitation number operator $\hat{N}=2\hat{\sigma}_{+}\hat{\sigma}_{-}+\hat{a}^{\dagger}\hat{a}$ is a conserved quantity based on the commutative relation $[\hat{N},\hat{H}_{\text{2pJC}}]=0$. The subspaces corresponding to the weighted excitation number $N=0, 1, 2, 3, \cdot\cdot\cdot, n, \cdot\cdot\cdot$ are spanned over the basis states $\{\vert g,0\rangle\}$, $\{\vert g,1\rangle\}$, $\{\vert g,2\rangle,\vert e,0\rangle\}$, $\{\vert g,3\rangle,\vert e,1\rangle\}$, $\cdot\cdot\cdot$, $\{\vert g,n\rangle,\vert e,n-2\rangle\}, \cdot\cdot\cdot$, where $\vert n\rangle$ $(n=2,3,4\cdot\cdot\cdot)$ denotes the number state of the cavity-field mode.

In the zero-excitation subspace, the eigen-equation can be obtained as $\hat{H}_{\text{2pJC}}\vert\varepsilon_{0}\rangle=\varepsilon_{0}\vert\varepsilon_{0}\rangle$ with the eigenstate $\vert\varepsilon_{0}\rangle=\vert g,0\rangle$ and the eigenvalue $\varepsilon_{0}=0$. In the one-excitation subspace, the eigenequation can be written as $\hat{H}_{\text{2pJC}}\vert\varepsilon_{1}\rangle=\varepsilon_{1}\vert\varepsilon_{1}\rangle$ with the eigenstate $\vert\varepsilon_{1}\rangle=\vert g,1\rangle$ and the eigenvalue $\varepsilon_{1}=\omega_{c}$. In the $n$-excitation $(n\geq2)$ subspace, the eigen-equation can be expressed as $\hat{H}_{\text{2pJC}}\vert\varepsilon_{n\pm}\rangle=\varepsilon_{n\pm}\vert\varepsilon_{n\pm}\rangle$, where the eigenvalues and eigenstates are, respectively, obtained as
\begin{equation}
\varepsilon_{n\pm}=\frac{2(n-1)\omega_{c}+\omega_{0}}{2}\pm\frac{\sqrt{(2\omega_{c}-\omega_{0})^{2}+4n(n-1)J^{2}}}{2},
\end{equation}
and
\begin{equation}\label{eigen}
\vert\varepsilon_{n\pm}\rangle=C_{g,n}^{[\pm]}\vert g,n\rangle+C_{e,n-2}^{[\pm]}\vert e,n-2\rangle.
\end{equation}
The superposition coefficients in Eq.~(\ref{eigen}) are given by
\begin{subequations}
\begin{align}
C_{g,n}^{[+]}&=C_{e,n-2}^{[-]}=\cos\theta_{n}, \\
C_{e,n-2}^{[+]}&=-C_{g,n}^{[-]}=\sin\theta_{n},
\end{align}
\end{subequations}
with the mixing angle $\theta_{n}$ defined by $\tan(2\theta_{n})=2\sqrt{n(n-1)}J/(2\omega_{c}-\omega_{0})$. In the resonant case $\omega_{0}=2\omega_{c}$, the eigenvalues and eigenstates of the system are reduced as $\varepsilon_{n\pm}=n\omega_{c}\pm\sqrt{n(n-1)}J$ and $\vert\varepsilon_{n\pm}\rangle=(\vert g,n\rangle\pm\vert e,n-2\rangle)/\sqrt{2}$, respectively.

To study the 1PB and 2PB effects, we consider the weak-driving case in which the Hilbert space of the cavity field can be truncated up to $n=4$. Figure~\ref{Fig1}(b) shows the eigenenergy spectrum of the Hamiltonian $\hat{H}_{\text{2pJC}}$ versus the atomic frequency $\omega_{0}$ in units of the cavity-field frequency $\omega_{c}$ in the subspace associated with zero, one, two, three, and four photons for $J/\omega_{c}=0.01$. Obviously, the eigenenergy spectrum of the system is anharmonic in the vicinity of the resonance point ($\omega_{0}\approx2\omega_{c}$), which means that the PB effect is more evident around the resonance point. In Fig.~\ref{Fig1}(c), we show the eigenenergy spectrum of the Hamiltonian $\hat{H}_{\text{2pJC}}$ in the resonant case $\omega_{0}=2\omega_{c}$. Below, we study the 1PB and 2PB effects in this system by driving either the cavity, $\hat{H}_{d}=\Omega(\hat{a}^{\dagger}e^{-i\omega_{d}t}+\hat{a}e^{i\omega_{d}t})$, or the atom, $\hat{H}_{d}^{\prime}=\Omega_{L}(\hat{\sigma}_{+}e^{-i\omega_{L}t}+\hat{\sigma}_{-}e^{i\omega_{L}t})$. Here $\Omega$ ($\Omega_{L}$) and $\omega_{d}$ ($\omega_{L}$) are the driving strength and driving frequency of the cavity field (atom), respectively. When the driving frequency $\omega_{d}$ matches the energy separation $\omega_{c}$ between the first excited state $\vert\varepsilon_{1}\rangle$ and the ground state $\vert\varepsilon_{0}\rangle$, the single-photon transition ($\vert\varepsilon_{0}\rangle\rightarrow\vert\varepsilon_{1}\rangle$) becomes resonant, but the subsequent transitions ($\vert\varepsilon_{1}\rangle\rightarrow\vert\varepsilon_{2\pm}\rangle$) induced by the second photon are blockaded due to the anharmonicity of the eigenenergy spectrum. This indicates that the 1PB effect can occur in this system. Similarly, when the driving frequency $2\omega_{d}$ ($\omega_{L}$) matches the energy-level differences $2\omega_{c}\pm\sqrt{2}J$ between $\vert\varepsilon_{2\pm}\rangle$ and $\vert\varepsilon_{0}\rangle$, the two-photon transitions ($\vert\varepsilon_{0}\rangle\rightarrow\vert\varepsilon_{2\pm}\rangle$) become resonant, while the subsequent transitions ($\vert\varepsilon_{2\pm}\rangle\rightarrow\vert\varepsilon_{3\pm}\rangle$) are blockaded, i.e., the 2PB effect can be observed in this system.

In the weak-driving case, the transition amplitude between two states is proportional to the ratio of the transition matrix element to the transition detuning. The transition behavior in this model induced by the driving terms can be analyzed by calculating the transition amplitudes between the involved energy levels in the eigenstate representation. In Figs.~\ref{Fig1}(d) and~\ref{Fig1}(e), we show the transition matrix elements between different energy levels in the resonant case $\omega_{0}=2\omega_{c}$ when the cavity field and the atom are driven~\cite{hamsen2017TwoPhoton}, respectively. In our following discussion, we analyze the locations of the peaks and dips in these correlation functions by calculating the resonance conditions for single- and multi-photon transitions.

\section{Criteria of the $n$PB and PIT effects\label{PBPIT}}

The physical picture of the $n$PB and PIT effects can be explained by analyzing the photon-number distribution $P_{n}\equiv\langle|n\rangle\langle n|\rangle$ and the equal-time $n$th-order correlation function $g^{(n)}(0)\equiv\langle\hat{a}^{\dagger n}\hat{a}^{n}\rangle/\langle\hat{n}\rangle^{n}$, with $\hat{n}=\hat{a}^{\dagger}\hat{a}$ being the photon number operator. In order to observe the $n$PB effect, Hamsen \textit{et al.}~\cite{hamsen2017TwoPhoton} proposed two criteria. The first criterion is based on a comparison between the photon-number distributions and the Poisson distributions of a coherent state. In this case, the criterion is defined by
\begin{equation}
\label{compar}
P_{n}\geq \mathcal{P}_{n}, \qquad
P_{m>n}<\mathcal{P}_{m>n},
\end{equation}
where $\mathcal{P}_{n}$ are the Poisson distributions defined by $
\mathcal{P}_{n}=\langle\hat{n}\rangle^{n}\exp{(-\langle\hat{n}\rangle)}/n!$,  with $\langle\hat{n}\rangle$ being the average photon number. Equation~(\ref{compar}) indicates that the probability of $n$ photons is enhanced and the probabilities of other photon numbers ($>n$) are suppressed for the $n$PB effect. The other criterion is based on the equal-time $n$th-order correlation function $g^{(n)}(0)$. In the case of weak driving, the mean photon number is very small, i.e., $\langle\hat{n}\rangle\ll1$. The criteria for the correlation functions for the $n$PB effect are~\cite{hamsen2017TwoPhoton,huang2018Nonreciprocal}
\begin{equation}
\label{nPB}
g^{(n)}(0)\geq1, \quad
g^{(n+1)}(0)<1,
\end{equation}
which means the $n$th-order super-Poissonian photon statistics or Poisson photon statistics, and the $(n+1)$th-order sub-Poissonian photon statistics. For instance, the correlation functions $g^{(2)}(0)\geq1$ and $g^{(3)}(0)<1$ are satisfied for the 2PB effect. The correlation function $g^{(2)}(0)\ll1$ is a signature of the 1PB effect.

On the other hand, for PIT, the absorption of the first photon favors that of the second or subsequent photons, so the PIT effect is usually characterized by the super-Poissonian photon statistics. Obviously, the process of PIT is inverse to the PB. Therefore, we refer to PIT if the $n$th-order correlation functions $g^{(n)}(0)>1$ ($n=2,3$) are satisfied in the weak-driving case~\cite{huang2018Nonreciprocal}. Note that the criteria for PIT have been analyzed in more detail in Refs.~\cite{faraon2008Coherent,majumdar2012LossEnabled,majumdar2012Probing,xu2013Photoninduced,rundquist2014Nonclassical,kowalewska-kudlaszyk2019Twophoton}.

It should be mentioned that the criteria for the $n$PB in Eq.~(\ref{nPB}) and PIT are mainly used for the single-photon physical transition process. In the two-photon JC model, the single-photon physical transition process occurs when the cavity field is driven, while the two-photon physical transition process, namely, the creation or annihilation of two photons, happens when the atom is driven. Hence, we propose that the criteria for the correlation functions for the $n$PB effect in the two-photon physical transition process should be
\begin{equation}
g^{(n)}(0)\geq1, \quad
g^{(n+1)}(0)<1, \quad
g^{(n+2)}(0)<1.
\end{equation}
For instance, in the atom-driving case, the correlation functions $g^{(2)}(0)\geq1$ and $g^{(n)}(0)<1$ ($n=3,4$) are satisfied for the 2PB effect, and the PIT effect can be characterized by the conditions of $g^{(n)}(0)>1$ ($n=2,3,4$).

\section{PB in the single-photon cavity-field-driving case \label{cavdrsec}}

In this section, we study the PB and PIT effects by analytically and numerically calculating the second- and third-order correlation functions for the cavity mode in the single-photon cavity-field-driving case.

\subsection{Analytical results}

When the cavity field is continuously driven by a monochromatic weak field, the driving Hamiltonian is described by
\begin{equation}
\hat{H}_{d}=\Omega(\hat{a}^{\dagger}e^{-i\omega_{d}t}+\hat{a}e^{i\omega_{d}t}),\label{Hdri}
\end{equation}
where $\Omega$ and $\omega_{d}$ are the driving strength and driving frequency, respectively. Then the total Hamiltonian of the system becomes $\hat{H}_{\mathrm{sys}}=\hat{H}_{\mathrm{2pJC}}+\hat{H}_{d}$. In a rotating frame defined by the unitary operator $\exp[-i\omega_{d}(\hat{a}^{\dagger}\hat{a}+\hat{\sigma}_{z})t]$, the Hamiltonian of the system becomes
\begin{eqnarray}
\hat{H}_{\text{sys}}^{(I)}&=&\Delta_{c}\hat{a}^{\dagger}\hat{a}+\Delta_{0}\hat{\sigma}_{+}\hat{\sigma}_{-}
+J(\hat{a}^{\dagger2}\hat{\sigma}_{-}+\hat{\sigma}_{+}\hat{a}^{2}) \nonumber\\
&&+\Omega(\hat{a}^{\dagger}+\hat{a}),\label{HsysInt}
\end{eqnarray}
where $\Delta_{c}=\omega_{c}-\omega_{d}$ ($\Delta_{0}=\omega_{0}-2\omega_{d}$) is the detuning of the cavity-field (atomic) frequency with respect to the driving frequency.

To include the influence of the dissipations of the cavity field and the atom on the PB effect, we phenomenologically add the imaginary dissipation terms into Hamiltonian~(\ref{HsysInt}) as
\begin{eqnarray}
\hat{H}_{\text{eff}}&=&(\Delta_{c}-i\kappa/2)\hat{a}^{\dagger}\hat{a}+(\Delta_{0}-i\gamma/2)\hat{\sigma}_{+}\hat{\sigma}_{-}\nonumber\\
&&+J(\hat{a}^{\dagger2}\hat{\sigma}_{-}+\hat{\sigma}_{+}\hat{a}^{2})+\Omega(\hat{a}^{\dagger }+\hat{a}),
\end{eqnarray}
where we have assumed that the cavity field and the atom are connected with two individual vacuum reservoirs, with $\kappa$ and $\gamma$ being the corresponding decay rates.

In the weak-driving regime ($\Omega\ll\kappa$), we truncate the Hilbert space of the cavity field up to $n=3$. In this subspace, a general state of the system can be written as
\begin{eqnarray}
\vert\psi(t)\rangle&=&C_{g0}(t)\vert g,0\rangle+C_{g1}(t)\vert g,1\rangle+C_{g2}(t)\vert g,2\rangle        \nonumber\\
&&+C_{e0}(t)\vert e,0\rangle+C_{g3}(t)\vert g,3\rangle+C_{e1}(t)\vert e,1\rangle,\quad
\end{eqnarray}
where the coefficients $C_{sj}(t)$ ($s=g,e$ and $j=0,1,2,3$) are the probability amplitudes. Based on the Schr\"{o}dinger equation $i\vert\dot{\psi}(t)\rangle=\hat{H}_{\mathrm{eff}}\vert\psi(t)\rangle$, we can obtain the equations of motion for these probability amplitudes $C_{sj}(t)$. Under the weak-driving condition ($\Omega\ll\kappa$), we have the approximate scales $C_{g0}\sim1$, $C_{g1}\sim\Omega/\kappa$, $\{C_{g2},C_{e0}\}\sim\Omega^{2}/\kappa^{2}$, and $\{C_{g3},C_{e1}\}\sim\Omega^{3}/\kappa^{3}$, i.e., $C_{g0}\gg C_{g1}\gg\{C_{g2},C_{e0}\}\gg\{C_{g3},C_{e1}\}$. The equations of motion for $C_{sj}(t)$ can be approximately solved by using a perturbation method, i.e., discard the higher-order terms in the equations of motion for the lower-order variables. We assume $C_{g0}(0)=1$, then the steady-state solutions of the probability amplitudes can be obtained by setting $\partial C_{sj}/\partial t=0$ as
\begin{subequations}
\label{stestasol}
\begin{align}
C_{g0}=&\ 1,   \\
C_{g1}=&-\frac{2\Omega}{2\Delta_{c}-i\kappa },   \\
C_{g2}=&\ 2\sqrt{2}i(\gamma+2i\Delta_{0})\Omega^{2}W^{-1},   \\
C_{e0}=&\ 8J\Omega^{2}W^{-1}, \\
C_{g3}=&-\frac{4\sqrt{6}[8J^{2}-(\gamma+2i\Delta_{0})V]\Omega^{3}}{3W[8J^{2}+(2i\Delta_{c}+\kappa)V]},    \\
C_{e1}=&-\frac{i16JV\Omega^{3}}{W[8J^{2}+(2i\Delta_{c}+\kappa)V]},
\end{align}
\end{subequations}
where we have introduced the variables
\begin{subequations}
\begin{align}
W&=(2\Delta_{c}-i\kappa)[4J^{2}+(\gamma+2i\Delta_{0})(2i\Delta_{c}+\kappa)],  \\
V&=\gamma+2i(\Delta_{0}+\Delta_{c})+\kappa.
\end{align}
\end{subequations}

Based on Eq.~(\ref{stestasol}), we obtain the steady state of the system, then the equal-time second- and third-order correlation functions can be expressed as
\begin{eqnarray}
g^{(2)}(0)\equiv\frac{\langle \hat{a}^{\dagger2}\hat{a}^{2} \rangle}{\langle \hat{a}^{\dagger}\hat{a} \rangle^2}=\frac{2P_{2}+6P_{3}}{(P_{1}+2P_{2}+3P_{3})^2}\approx\frac{2P_{2}}{P_{1}^2}, \label{analyticalg2}
\end{eqnarray}
and
\begin{eqnarray}
g^{(3)}(0)\equiv\frac{\langle \hat{a}^{\dagger3}\hat{a}^{3} \rangle}{\langle \hat{a}^{\dagger}\hat{a} \rangle^3}=\frac{6P_{3}}{(P_{1}+2P_{2}+3P_{3})^3}\approx\frac{6P_{3}}{P_{1}^3}, \label{analyticalg3}
\end{eqnarray}
where the photon-number distributions are given by
\begin{eqnarray}
\label{phnumpro}
P_{n=0,1}&=&(\vert C_{g,n}\vert^{2}+\vert C_{e,n}\vert^{2})/\mathcal{N}, \nonumber\\
P_{m=2,3}&=&\vert C_{g,m}\vert^{2}/\mathcal{N},
\end{eqnarray}
with the normalization constant $\mathcal{N}=\vert C_{g0}\vert^{2}+\vert C_{g1}\vert^{2}+\vert C_{g2}\vert^{2}+\vert C_{e0}\vert^{2}+\vert C_{g3}\vert^{2}+\vert C_{e1}\vert^{2}$. For the weak-driving case, this normalization constant can be omitted because $\mathcal{N}\approx1$.

In this low-excitation subspace, a case corresponding to a perfect photon blockade is $C_{g2}=0$, which means that there are no two-photon probabilities in the cavity. The parameter condition for this perfect 1PB effect can be obtained as
\begin{equation}
\Delta_{0}=0,\hspace{0.5 cm}\gamma=0.
\end{equation}
The optimal parameter condition can be explained based on the destructive quantum interference between the two paths ($\vert\varepsilon_{0}\rangle\rightarrow\vert\varepsilon_{2\pm}\rangle$)~\cite{zou2018Photon}. A detailed analysis of the destructive quantum interference is given in the next section.

\subsection{Numerical results}

In order to confirm our analytical results, we numerically calculate the equal-time second- and third-order correlation functions of the cavity mode. Numerical computations were performed using the Python package QuTiP~\cite{johansson2012QuTiP,johansson2013QuTiP}. We assume that the cavity and the two-level atom are connected with two individual vacuum baths. Then the dynamics of the system is governed by the quantum master equation
\begin{eqnarray}
\frac{d\hat{\rho}(t)}{dt}&=&i[\hat{\rho}(t),\hat{H}_{\text{sys}}^{(I)}]+\frac{\kappa}{2}[2\hat{a}\hat{\rho}(t)\hat{a}^{\dagger}-\hat{a}^{\dagger}\hat{a}\hat{\rho}(t)-\hat{\rho}(t)\hat{a}^{\dagger}\hat{a}]\nonumber\\
&&+\frac{\gamma}{2}[2\hat{\sigma}_{-}\hat{\rho}(t)\hat{\sigma}_{+}-\hat{\sigma}_{+}\hat{\sigma}_{-}\hat{\rho}(t)-\hat{\rho}(t)\hat{\sigma}_{+}\hat{\sigma}_{-}],\label{SME}
\end{eqnarray}
where $\kappa$ ($\gamma$) is the decay rate of the cavity (atom). By numerically solving Eq.~(\ref{SME}), we can get the steady-state density operator $\hat{\rho}_{\mathrm{ss}}$ of the system, and then the photon-number distributions $P_{n}=\mathrm{Tr}[\vert n\rangle \langle n\vert \hat{\rho}_{\mathrm{ss}}]$ can be calculated. The equal-time $n$th-order correlation functions can also be obtained by $g^{(n)}(0)=\mathrm{Tr}(\hat{a}^{\dagger n}\hat{a}^{n}\hat{\rho}_{\mathrm{ss}})/[\mathrm{Tr}(\hat{a}^{\dagger}\hat{a}\hat{\rho}_{\mathrm{ss}})]^n$.

\begin{figure}
\center
\includegraphics[width=0.47 \textwidth]{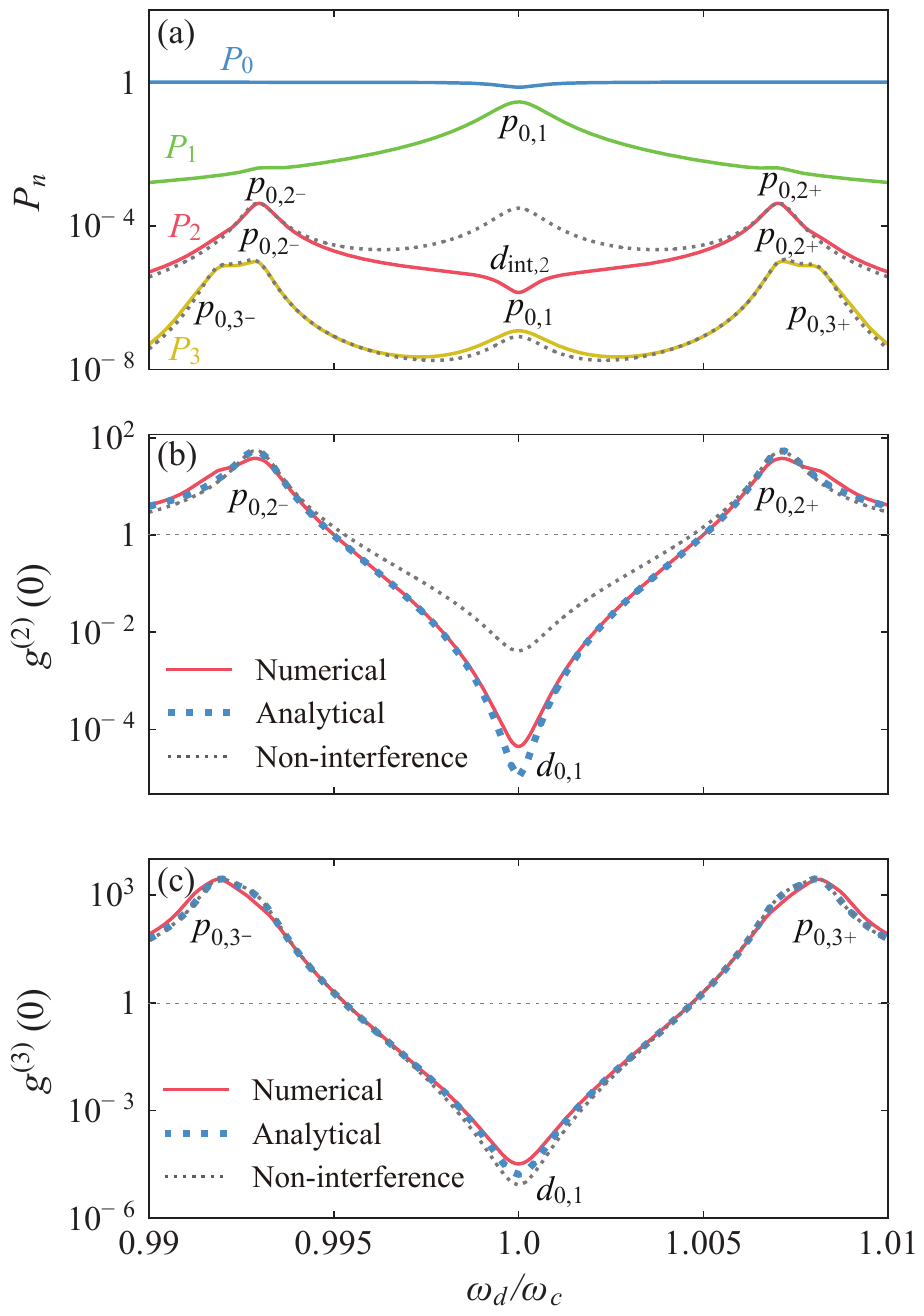}
\caption{(Color online) (a) Photon-number distributions $P_{n=0,1,2,3}$ versus cavity-field driving frequency $\omega_{d}/\omega_{c}$ in the resonant case $\omega_{0}/\omega_{c}=2$. Solid colored curves and dotted gray curves are plotted based on the analytical and non-interference results, respectively. Second- and third-order correlation functions (b) $g^{(2)}(0)$ and (c) $g^{(3)}(0)$ versus the cavity-field driving frequency $\omega_{d}/\omega_{c}$. Solid red (dotted blue) curves represent the numerical (analytical) results. Other parameters used are $J/\omega_{c}=0.01$, $\kappa/\omega_{c}=\gamma/\omega_{c}=0.001$, and $\Omega/\kappa=0.4$.}
\label{Fig2}
\end{figure}

For studying the PB effect in this model, we consider both the resonant ($\omega_{0}=2\omega_{c}$) and the off-resonant ($\omega_{0}\neq2\omega_{c}$) cases. In Fig.~\ref{Fig2}(a), we plot the photon-number distributions $P_{n=0,1,2,3}$ as functions of the cavity-field driving frequency $\omega_{d}/\omega_{c}$ in the resonant case $\omega_{0}=2\omega_{c}$. We see the relations $P_{0}\approx1$ and $P_{0}\gg P_{1}\gg P_{2}\gg P_{3}$ in the weak-driving case. In addition, there is a peak located at $\omega_{d}/\omega_{c}=1$ in the curve of $P_{1}$ (solid green curve), while there are a dip and two peaks in the curve of $P_{2}$ (solid red curve), with the locations $\omega_{d}/\omega_{c}=1$ and $1\pm J/(\sqrt{2}\omega_{c})$, respectively. By analyzing the energy spectrum of this system, we find that the locations of these peaks in the curves of $P_{1}$ and $P_{2}$ are determined by the single- and two-photon resonance transitions $\vert\varepsilon_{0}\rangle\rightarrow\vert\varepsilon_{1}\rangle$ and $\vert\varepsilon_{0}\rangle\rightarrow\vert\varepsilon_{2\pm}\rangle$, respectively. To be clearer, we mark these peaks in the curves of $P_{1}$ and $P_{2}$ as $p_{0,1}$ and $p_{0,2\pm}$. In the curve of $P_{3}$ (solid yellow curve), we see that there are five peaks located at $\omega_{d}/\omega_{c}=1$, $1\pm J/(\sqrt{2}\omega_{c})$, and $1\pm \sqrt{6}J/(3\omega_{c})$, respectively. The locations of the two peaks $p_{0,3\pm}$ are determined by the three-photon resonance transitions $\vert\varepsilon_{0}\rangle\rightarrow\vert\varepsilon_{3\pm}\rangle$, while the other three peaks are induced by the single- and two-photon resonance transitions, and hence the locations of the three peaks are the same as those of the three peaks in the curves of $P_{1}$ and $P_{2}$. The dip in the curve of $P_{2}$ can be explained by the destructive quantum interference between the two different paths ($\vert\varepsilon_{0}\rangle\rightarrow\vert\varepsilon_{2\pm}\rangle$) of the two-photon excitation. Thus, we label this dip $d_{\text{int},2}$. To prove this point, in the following we present a detailed analysis of the influence of the quantum interference effect in the eigenstate representation on the photon-number distributions.

In the eigenstate representation, a general pure state of the system in the low-excitation subspace can be expressed as
\begin{eqnarray}
\vert\Psi(t)\rangle&=&D_{0}(t)\vert\varepsilon_{0}\rangle+D_{1}(t)\vert\varepsilon_{1}\rangle+D_{2-}(t)\vert\varepsilon_{2-}\rangle \nonumber\\
&&+D_{2+}(t)\vert\varepsilon_{2+}\rangle+D_{3-}(t)\vert\varepsilon_{3-}\rangle+D_{3+}(t)\vert\varepsilon_{3+}\rangle.\qquad
\end{eqnarray}
According to the Schr\"{o}dinger equation $i\vert\dot{\Psi}(t)\rangle=\hat{H}_{\text{eff}}\vert\Psi(t)\rangle$, we can obtain the equations of motion for these probability amplitudes $D_{i}(t)$ ($i=0,1$) and $D_{js}(t)$ ($j=2,3$ and $s=\pm$). The steady-state solutions of these probability amplitudes can be obtained by using the perturbation method. The zero-photon (one-photon) probability can be expressed as $P_{0}\approx\vert D_{0}\vert^{2}$ ($P_{1}\approx\vert D_{1}\vert^{2}$) because $C_{g0}\gg C_{e0}$ ($C_{g1}\gg C_{e1}$). The two- and three-photon probabilities can also be obtained as
\begin{subequations}
\label{interfer}
\begin{align}
P_{2}=&\left\vert D_{2-}C_{g2}^{[-]}\right\vert^{2}+\left\vert D_{2+}C_{g2}^{[+]}\right\vert^{2}+D_{2-}^{\ast}D_{2+}C_{g2}^{[-]\ast}C_{g2}^{[+]} \nonumber\\ &+D_{2+}^{\ast}D_{2-}C_{g2}^{[+]\ast}C_{g2}^{[-]}, \label{interfera} \\
P_{3}=&\left\vert D_{3-}C_{g3}^{[-]}\right\vert^{2}+\left\vert D_{3+}C_{g3}^{[+]}\right\vert^{2}+D_{3-}^{\ast}D_{3+}C_{g3}^{[-]\ast}C_{g3}^{[+]} \nonumber\\ &+D_{3+}^{\ast}D_{3-}C_{g3}^{[+]\ast}C_{g3}^{[-]}, \label{interferb}
\end{align}
\end{subequations}
where the first two terms in Eq.~(\ref{interfera}) are the two-photon probability of the non-quantum-interference contribution, and the remanining terms (cross terms) are induced by quantum interference between the two different paths of the two-photon excitation. To confirm the quantum interference effect, we show the non-quantum-interference part (dotted gray curve) of $P_{2}$ as a reference in Fig.~\ref{Fig2}(a). Here we see that the two peaks in the curve of $P_{2}$ show an excellent agreement with those of the non-quantum-interference result, while the dip $d_{\text{int},2}$ in the curve of $P_{2}$ becomes a peak in the non-quantum-interference result. Therefore, the dip in the curve of $P_{2}$ can be explained based on the destructive quantum interference between the two paths ($\vert\varepsilon_{0}\rangle\rightarrow\vert\varepsilon_{2\pm}\rangle$).
\begin{figure}
\center
\includegraphics[width=0.47 \textwidth]{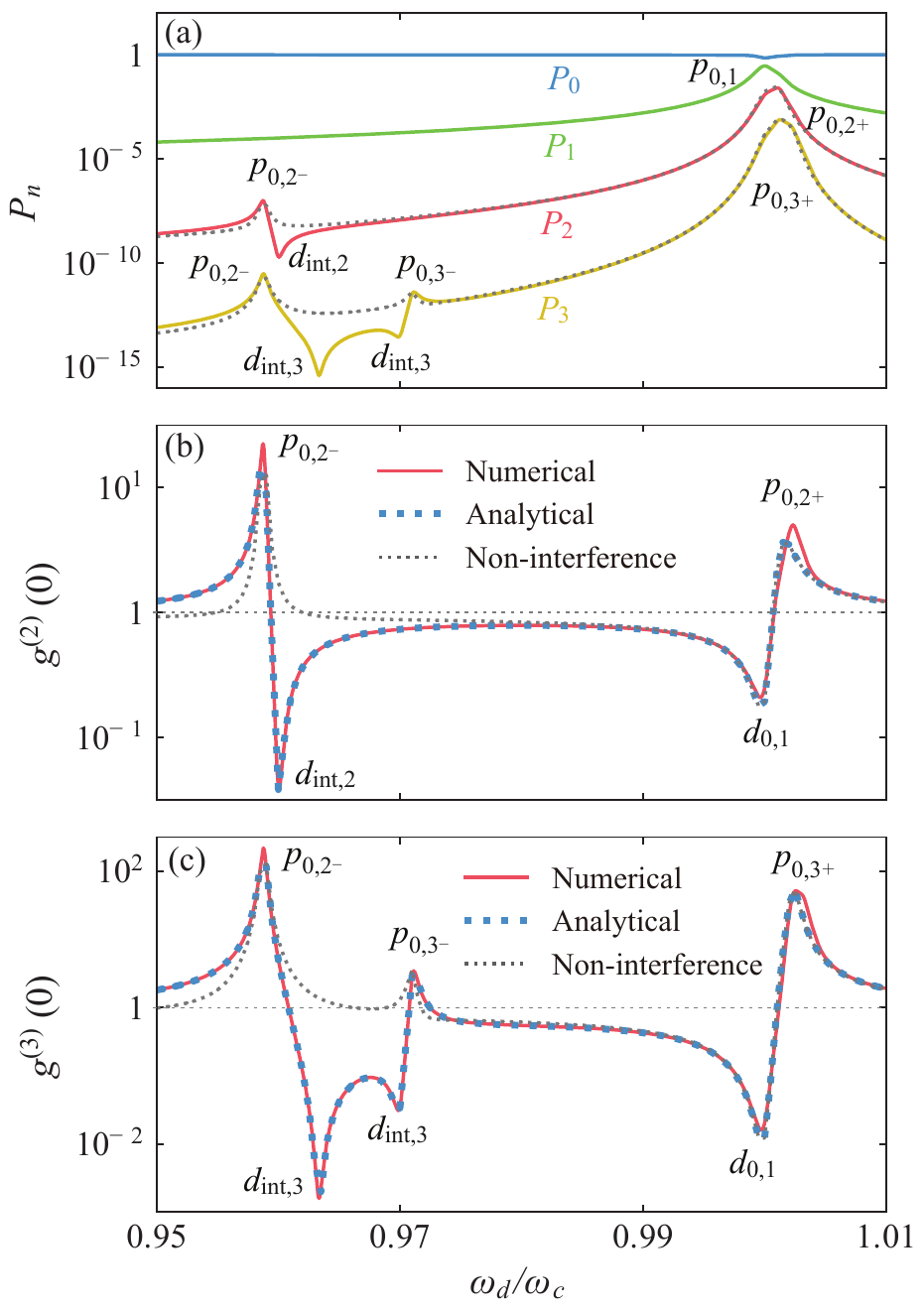}
\caption{(Color online) (a) Photon-number distributions $P_{n=0,1,2,3}$ versus cavity-field driving frequency $\omega_{d}/\omega_{c}$ in the off-resonant case $\omega_{0}/\omega_{c}=1.92$. Solid colored curves and dotted gray curves are plotted based on the analytical and non-interference results, respectively. The second- and third-order correlation function (b) $g^{(2)}(0)$ and (c) $g^{(3)}(0)$ versus the cavity-field driving frequency $\omega_{d}/\omega_{c}$. Solid red (dotted blue) curves represent the numerical (analytical) results. Other parameters used are the same as given in Fig.~\ref{Fig2}.}
\label{Fig3}
\end{figure}

To seek the optimal cavity-field driving frequency of the 1PB, in Fig.~\ref{Fig2}(b) we plot the second-order correlation function $g^{(2)}(0)$ versus $\omega_{d}/\omega_{c}$. In Fig.~\ref{Fig2}(b), we see that the analytical result shows an excellent agreement with the numerical result and that the two peaks of the non-quantum-interference result can also match well the analytical and numerical results, but the dip cannot. In addition, we find that the locations of the dip $d_{0,1}$ and the two peaks $p_{0,2\pm}$ in the curve of $g^{(2)}(0)$ correspond to single- and two-photon resonance transitions, respectively. In the single-photon resonance case, the 1PB effect can be observed because $g^{(2)}(0)\ll1$. In the two-photon resonance case, we see that $g^{(2)}(0)>1$. To further investigate the 2PB effect, we show the third-order correlation function $g^{(3)}(0)$ versus $\omega_{d}/\omega_{c}$ in Fig.~\ref{Fig2}(c). According to the expression $g^{(3)}(0)\approx6P_{3}/P_{1}^{3}$, we find that the locations of the dip ($d_{0,1}$) and the two peaks ($p_{0,3\pm}$) in the curve of $g^{(3)}(0)$ correspond to the single- and three-photon resonance transitions, respectively. In particular, the correlation functions exhibit $g^{(2)}(0)>1$ and $g^{(3)}(0)>1$ at $1\pm J/(\sqrt{2}\omega_{c})$, which is a signature of PIT in the two-photon resonance case.

Figure~\ref{Fig3}(a) displays $P_{n=0,1,2,3}$ as functions of $\omega_{d}/\omega_{c}$ in the off-resonant case $\omega_{0}=1.92\omega_{c}$. Here we can see that there is a peak $p_{0,1}$ in the curve of $P_{1}$ located at $\omega_{d}/\omega_{c}=1$. In addition, there are two peaks in the curve of $P_{2}$ located at $[3.92\pm\sqrt{0.0064+8(J/\omega_{c})^{2}}]/4$. The locations of the two main peaks $p_{0,2\pm}$ correspond to the two-photon resonance transitions $\vert\varepsilon_{0}\rangle\rightarrow\vert\varepsilon_{2\pm}\rangle$. We point out that there is a dip $d_{\text{int},2}$ located at $\omega_{d}/\omega_{c}=0.96$ in the curve of $P_{2}$, which disappears in the non-quantum-interference result (dotted gray curve). Here, the location of the dip in the curve of $P_{2}$ is different from that of the peak in the curve of $P_{1}$, differently from the results in the resonant case $\omega_{0}=2\omega_{c}$. In the curve of $P_{3}$ there are three peaks, located at $[3.92-\sqrt{0.0064+8(J/\omega_{c})^{2}}]/4$ and $[5.92\pm\sqrt{0.0064+24(J/\omega_{c})^{2}}]/6$, respectively. The locations of these peaks ($p_{0,2-}$ and $p_{0,3\pm}$) match those of the two- and three-photon resonance transitions. Moreover, the two dips $d_{\text{int},3}$ in the curve of $P_{3}$ are induced by destructive quantum interference between the two transition paths ($\vert\varepsilon_{0}\rangle\rightarrow\vert\varepsilon_{3\pm}\rangle$) of the three-photon excitation, which can be confirmed by comprising the analytical result with the non-quantum-interference result (dotted gray curve).
\begin{figure}
\center
\includegraphics[width=0.45 \textwidth]{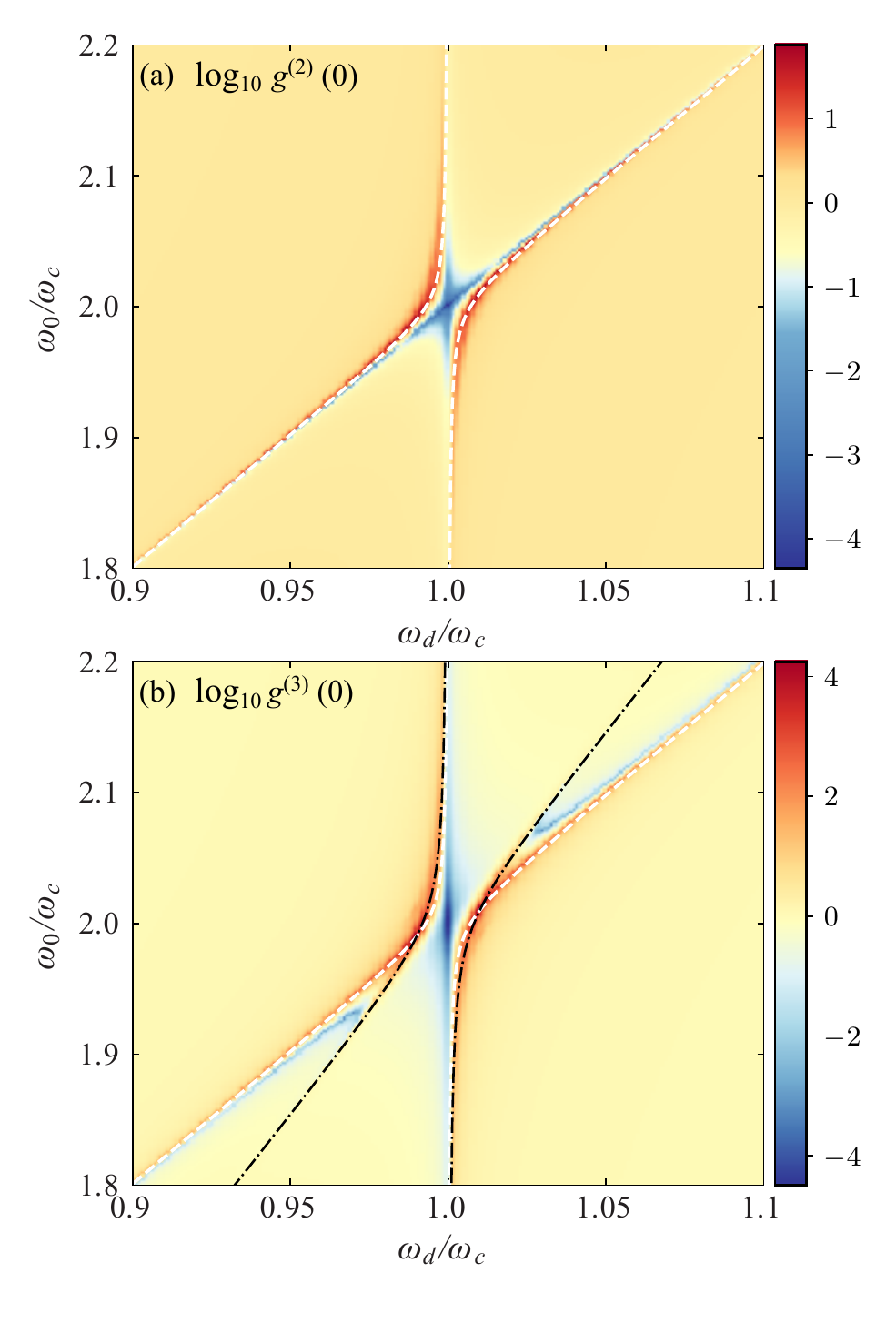}
\caption{(Color online) Plot of (a) $\log_{10}g^{(2)}(0)$ and (b) $\log_{10}g^{(3)}(0)$ as functions of $\omega_{d}/\omega_{c}$ and $\omega_{0}/\omega_{c}$. Dashed white curves and dash-dotted black curves correspond to the two- and three-photon resonance transitions, respectively. Other parameters used are the same as given in Fig.~\ref{Fig2}.}
\label{Fig4}
\end{figure}

In the off-resonant case $\omega_{0}=1.92\omega_{c}$, we analyze the optimal cavity-field driving frequency of 1PB by showing $g^{(2)}(0)$ as a function of $\omega_{d}/\omega_{c}$ in Fig.~\ref{Fig3}(b). It is found that there are two dips in the curve of $g^{(2)}(0)$, which is a signature of the 1PB effect. One of the two dips $d_{0,1}$ corresponds to the single-photon resonance transition; the other dip $d_{\text{int},2}$ is caused by destructive quantum interference between the two paths $\vert\varepsilon_{0}\rangle\rightarrow\vert\varepsilon_{2\pm}\rangle$. To further explain the quantum interference effect, we show the analytical result of the non-quantum-interference part (dotted gray curve). We find that the dip $d_{\text{int},2}$ caused by the quantum-interference effect disappears in the non-quantum-interference result. In Fig.~\ref{Fig3}(c), $g^{(3)}(0)$ is plotted as a function of $\omega_{d}/\omega_{c}$. We find that the locations of the three peaks ($p_{0,2-}$ and $p_{0,3\pm}$) in the curve of $g^{(3)}(0)$ correspond to two- and three-photon resonance transitions, respectively. However, the three dips ($d_{0,1}$ and two $d_{\text{int},3}$) in the curve of $g^{(3)}(0)$ are caused by the single-photon resonance transition and the destructive quantum interference between the two paths $\vert\varepsilon_{0}\rangle\rightarrow\vert\varepsilon_{3\pm}\rangle$, respectively. Moreover, the relations $g^{(2)}(0)>1$ and $g^{(3)}(0)>1$ indicate that PIT can be observed in the two-photon resonance case.
\begin{figure}
\center
\includegraphics[width=0.47 \textwidth]{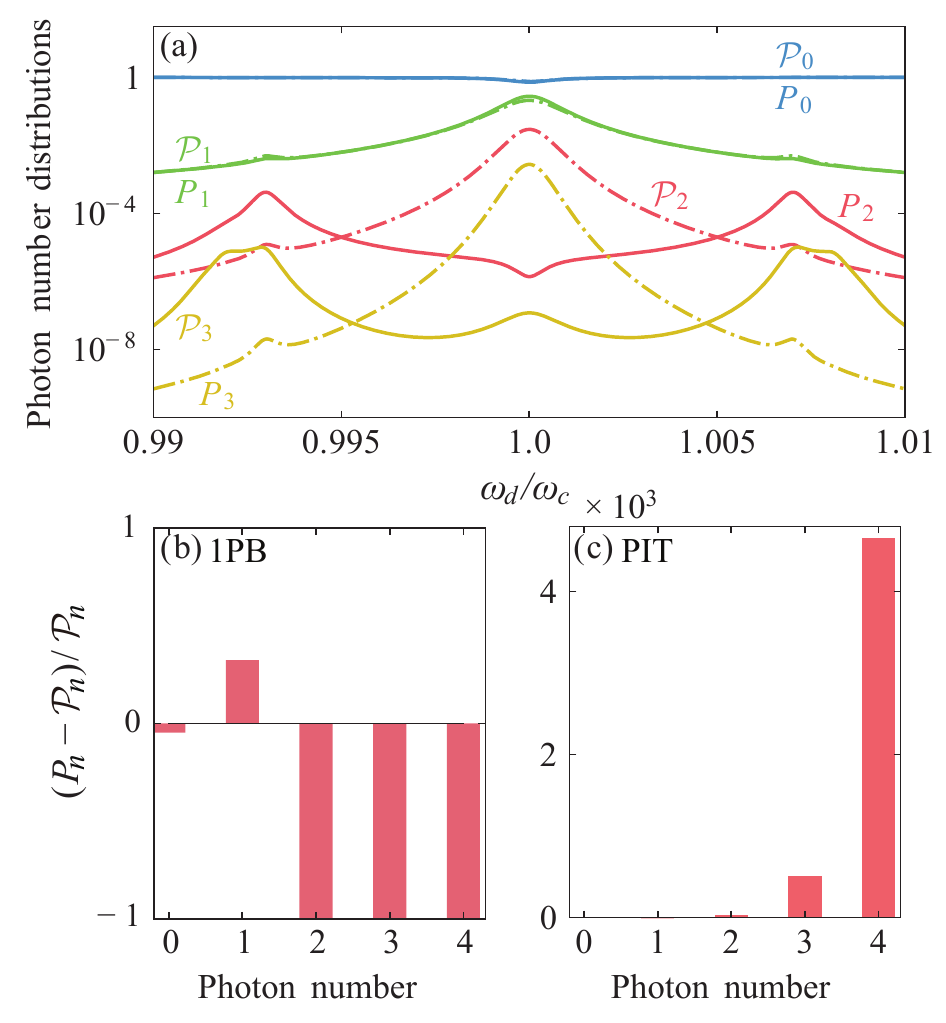}
\caption{(Color online) (a) Photon-number distributions $P_{n=0,1,2,3}$ (solid colored curves) and Poisson distributions $\mathcal{P}_{n=0,1,2,3}$ (dash-dotted colored curves) versus the driving frequency $\omega_{d}/\omega_{c}$ at $\omega_{0}/\omega_{c}=2$. Relative deviations of the photon-number distribution from the standard Poisson distribution with the same mean photon number located at (b) $\omega_{d}/\omega_{c}=1$ and (c) $\omega_{d}/\omega_{c}=1\pm J/\sqrt{2}\omega_{c}$. Other parameters used are the same as given in Fig.~\ref{Fig2}.}
\label{Fig5}
\end{figure}

With regard to the analysis of the off-resonant case, we only consider a particular case in Fig.~\ref{Fig3}. A more comprehensive analysis of the off-resonant case is shown in Fig.~\ref{Fig4}. We show $\log_{10}g^{(2)}(0)$ as a function of $\omega_{d}/\omega_{c}$ and $\omega_{0}/\omega_{c}$ in Fig.~\ref{Fig4}(a). It is clear that the optimal parameter conditions to observe the 1PB effect are $\omega_{d}/\omega_{c}=1$ and $\omega_{0}/\omega_{d}=2$, respectively, i.e., $\Delta_{c}=0$ and $\Delta_{0}=0$. The condition $\Delta_{c}=0$ can be explained based on the single-photon resonance transition $\vert\varepsilon_{0}\rangle\rightarrow\vert\varepsilon_{1}\rangle$, and the condition $\Delta_{0}=0$ can be interpreted by $C_{g2}=0$ corresponding to destructive quantum interference between the two paths $\vert\varepsilon_{0}\rangle\rightarrow\vert\varepsilon_{2\pm}\rangle$. We also observe in Fig.~\ref{Fig4}(a) that $g^{(2)}(0)>1$ in the two-photon resonance case. In order to further investigate PIT in the off-resonant case, $\log_{10}g^{(3)}(0)$ is plotted as a function of $\omega_{d}/\omega_{c}$ and $\omega_{0}/\omega_{c}$ in Fig.~\ref{Fig4}(b). Obviously, we see that $g^{(3)}(0)>1$ under the two-photon resonance transitions. Therefore, PIT can be observed in the two-photon resonance case because $g^{(2)}(0)>1$ and $g^{(3)}(0)>1$. This implies that the 1PB effect and PIT can occur by driving the cavity, while the 2PB effect cannot occur in this case.
\begin{figure*}
\center
\includegraphics[width=0.9 \textwidth]{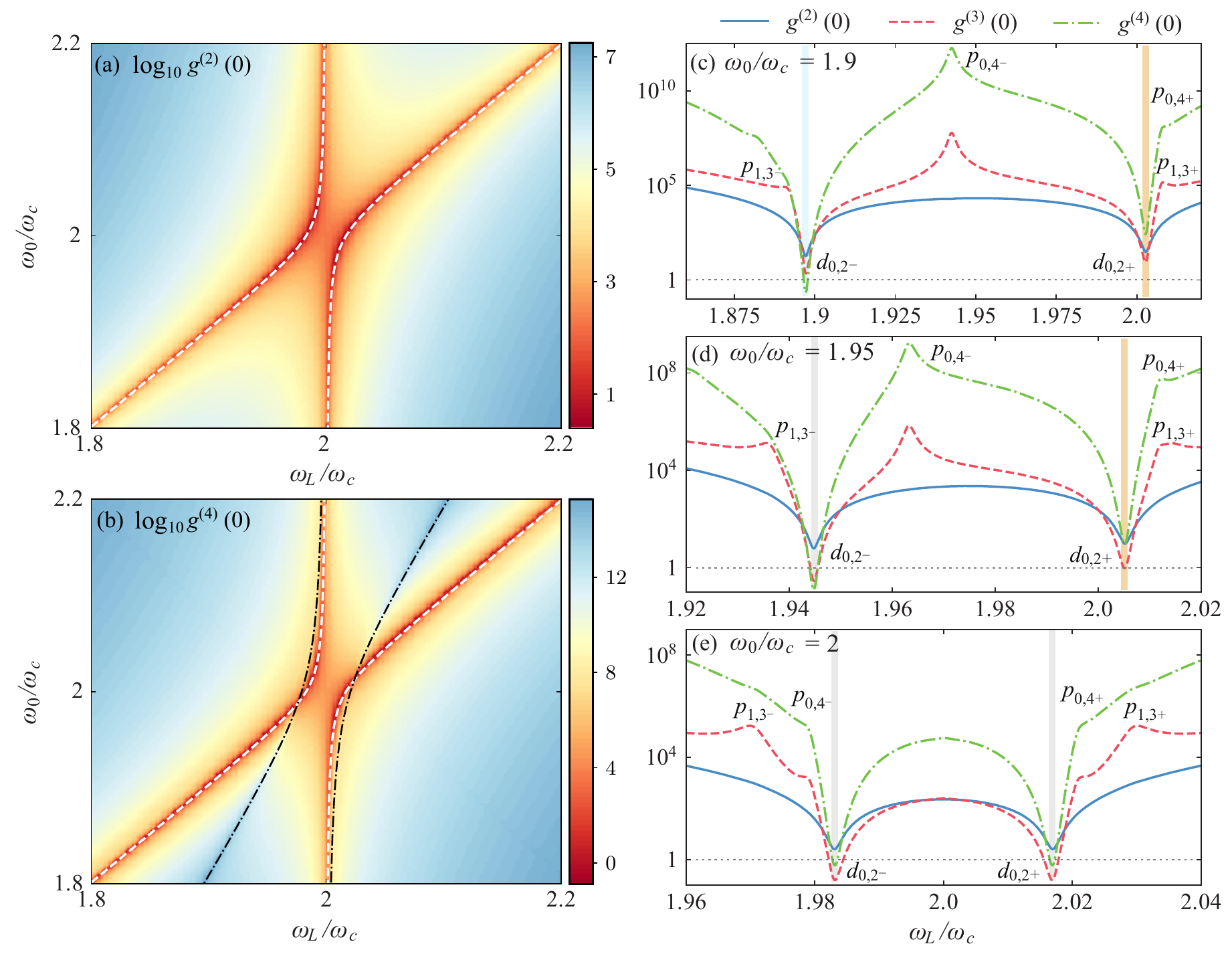}
\caption{(Color online) Plot of (a) $\log_{10}g^{(2)}(0)$ and (b) $\log_{10}g^{(4)}(0)$ as functions of $\omega_{L}/\omega_{c}$ and $\omega_{0}/\omega_{c}$. Dashed white curves and dash-dotted black curves correspond to the two- and four-photon resonance transitions, respectively. Correlation functions $g^{(2)}(0)$ (solid blue curves), $g^{(3)}(0)$ (dashed red curves), and $g^{(4)}(0)$ (dash-dotted green curves) as functions of $\omega_{L}/\omega_{c}$ at (c) $\omega_{0}/\omega_{c}=1.9$,  (d) $\omega_{0}/\omega_{c}=1.95$, and (e) $\omega_{0}/\omega_{c}=2$. Other parameters used are $J/\omega_{c}=0.012$, $\kappa/\omega_{c}=\gamma/\omega_{c}=0.001$, and $\Omega_{L}/\kappa=0.4$.}
\label{Fig6}
\end{figure*}

Our results can also be confirmed by comparing the photon-number distributions and the Poisson distributions. In Fig.~\ref{Fig5}(a) we plot $P_{n=0,1,2,3}$ (solid colored curves) and $\mathcal{P}_{n=0,1,2,3}$ (dash-dotted colored curves) as functions of  $\omega_{d}/\omega_{c}$ in the resonant case $\omega_{0}/\omega_{c}=2$. At the location of $\omega_{d}/\omega_{c}=1$ ($\omega_{d}/\omega_{c}=1\pm J/\sqrt{2}\omega_{c}$), the single-photon probability is enhanced (suppressed) because $P_{1}>\mathcal{P}_{1}$ ($P_{1}<\mathcal{P}_{1}$), while the two- and three-photon probabilities are suppressed (enhanced) because $P_{n}<\mathcal{P}_{n}$ ($P_{n}>\mathcal{P}_{n}$) for $n=2,3$. This means that the 1PB (PIT) effect occurs by driving the cavity in the single-photon (two-photon) resonance case. To further illustrate the 1PB and PIT effects, we show the relative deviations of the photon-number distribution to the standard Poisson distribution with the same mean photon number at $\omega_{d}/\omega_{c}=1$ in Fig.~\ref{Fig5}(b) and $\omega_{d}/\omega_{c}=1\pm J/\sqrt{2}\omega_{c}$ in Fig.~\ref{Fig5}(c). We observe in Fig.~\ref{Fig5}(b) that only the value of the single-photon relative population is greater than $0$, i.e., $P_{1}>\mathcal{P}_{1}$, which implies that the 1PB effect occurs. One can see in Fig.~\ref{Fig5}(c) that the relative population grows as the photon number increases, which is another signature of PIT.

We also investigate the influence of the coupling strength $J$ and the cavity-field decay rate $\kappa$ on the 1PB effect, and find that the correlation function $g^{(2)}(0)$ decreases (increases) monotonically as $J$ ($\kappa$) increases. This implies that the coupling strength (the cavity-field decay rate) enhances (attenuates) the 1PB effect.

\section{PB in the atom-driving case \label{atdrsec}}

In this section, we study PB effect in the atom-driving case by numerically calculating the correlation functions of the cavity-field mode.

\subsection{Theoretical analysis}
When a monochromatic weak driving field is applied to the atom, the driving Hamiltonian is described by
\begin{equation}
\hat{H}_{d}^{\prime}=\Omega_{L}(\hat{\sigma}_{+}e^{-i\omega_{L}t}+\hat{\sigma}_{-}e^{i\omega_{L}t}),
\end{equation}
where $\Omega_{L}$ and $\omega_{L}$ are the driving strength and driving frequency, respectively. In this case, the total Hamiltonian of the system reads $\hat{H}_{\text{sys}}^{\prime}=\hat{H}_{\text{2pJC}}+\hat{H}_{d}^{\prime}$. In a rotating frame defined by the unitary operator $\exp[-i\omega_{L}t(\hat{a}^{\dagger}\hat{a}+\hat{\sigma}_{z})/2]$, the Hamiltonian of the system becomes
\begin{eqnarray}
\hat{H}_{\text{sys}}^{\prime(I)}&=&\Delta^{\prime}_{c}\hat{a}^{\dagger}\hat{a}+\Delta^{\prime}_{0}\hat{\sigma}_{+}\hat{\sigma}_{-}+J(\hat{a}^{\dagger2}\hat{\sigma}_{-}+\hat{\sigma}_{+}\hat{a}^{2})\nonumber\\
&&+\Omega_{L}(\hat{\sigma}_{+}+\hat{\sigma}_{-}),
\end{eqnarray}
where $\Delta^{\prime}_{c}=\omega_{c}-\omega_{L}/2$ ($\Delta^{\prime}_{0}=\omega_{0}-\omega_{L}$) is the detuning of the cavity-field (atomic) frequency with respect to the driving frequency.

By numerically solving quantum master equation~(\ref{SME}) under the replacement of $\hat{H}_{\mathrm{sys}}^{(I)}\rightarrow\hat{H}_{\mathrm{sys}}^{\prime(I)}$, the steady-state density operator $\hat{\rho}_{\mathrm{ss}}^{\prime}$ of the system can be obtained and then we can calculate the photon-number distributions $P_{n}=\mathrm{Tr}[\vert n\rangle \langle n\vert \hat{\rho}_{\mathrm{ss}}^{\prime}]$ in the cavity-field mode. Similarly, the equal-time $n$th-order correlation functions can be obtained as
$g^{(n)}(0)=\mathrm{Tr}(\hat{a}^{\dagger n}\hat{a}^{n}\hat{\rho}_{\mathrm{ss}}^{\prime})/[\mathrm{Tr}(\hat{a}^{\dagger}\hat{a}\hat{\rho}_{\mathrm{ss}}^{\prime})]^n$. By analyzing these correlation functions, we can study the PB effect of this system.

\subsection{Numerical results}

When the atom is driven, the 1PB effect cannot be observed because the transition $\vert\varepsilon_{0}\rangle\stackrel{\Omega_{L}}{\longrightarrow}\vert\varepsilon_{1}\rangle$ is forbidden. In order to prove this, we show $\log_{10}g^{(2)}(0)$ as a function of $\omega_{L}/\omega_{c}$ and $\omega_{0}/\omega_{c}$ in Fig.~\ref{Fig6}(a). Clearly, we observe that $g^{(2)}(0)>1$ in the entire parameter area, which implies that the 1PB effect cannot appear by driving the atom. Differently from the cavity-field-driving case, two photons can be produced when driving the atom. To further study the 2PB effect, $\log_{10}g^{(4)}(0)$ is plotted as a function of $\omega_{L}/\omega_{c}$ and $\omega_{0}/\omega_{c}$ in Fig.~\ref{Fig6}(b). The dashed white (dash-dotted black) curves correspond to the two-photon (four-photon) resonance transitions, namely, $\vert\varepsilon_{0}\rangle\stackrel{\Omega_{L}}{\longrightarrow}\vert\varepsilon_{2\pm}\rangle$ ($\vert\varepsilon_{0}\rangle\stackrel{\Omega_{L}}{\longrightarrow}\vert\varepsilon_{4\pm}\rangle$). At the two-photon resonance transitions, the correlation function $g^{(4)}(0)<1$ for some parameters. This means that the 2PB can be observed by driving the atom in the two-photon resonance case, i.e., $g^{(2)}(0)>1$ and $g^{(4)}(0)<1$. To see this more clearly, in Figs.~\ref{Fig6}(c-e) the correlation functions $g^{(2)}(0)$ (solid blue curves), $g^{(3)}(0)$ (dashed red curves), and $g^{(4)}(0)$ (dash-dotted green curves) are plotted versus $\omega_{L}/\omega_{c}$ at different values of $\omega_{0}/\omega_{c}$. Here we find that the locations of these dips $d_{0,2\pm}$ in the curves of $g^{(2)}(0)$ correspond to two-photon resonance transitions, while the locations of these dips $d_{0,2\pm}$ and peaks ($p_{1,3\pm}$ and $p_{0,4\pm}$) in the curves of $g^{(3)}(0)$ correspond to two-, three-, and four-photon resonance transitions, respectively. In the curves of $g^{(4)}(0)$, the locations of these dips $d_{0,2\pm}$ and peaks $p_{0,4\pm}$ correspond to two- and four-photon resonance transitions, respectively. In Figs.~\ref{Fig6}(c-e), we see that the 2PB effect can occur in the gray areas because $g^{(2)}(0)>1$ and $g^{(n)}(0)<1$ ($n=3,4$). The yellow areas in Figs.~\ref{Fig6}(c) and~\ref{Fig6}(d) correspond to PIT due to $g^{(n)}(0)>1$ ($n=2,3,4$). It is noteworthy that the blue area in Fig.~\ref{Fig6}(c) indicates the enhanced two- and three-photon correlations [$g^{(n)}(0)>1$; $n=2,3$] and the suppressed four-photon correlation [$g^{(4)}(0)<1$].

\begin{figure}
\center
\includegraphics[width=0.47 \textwidth]{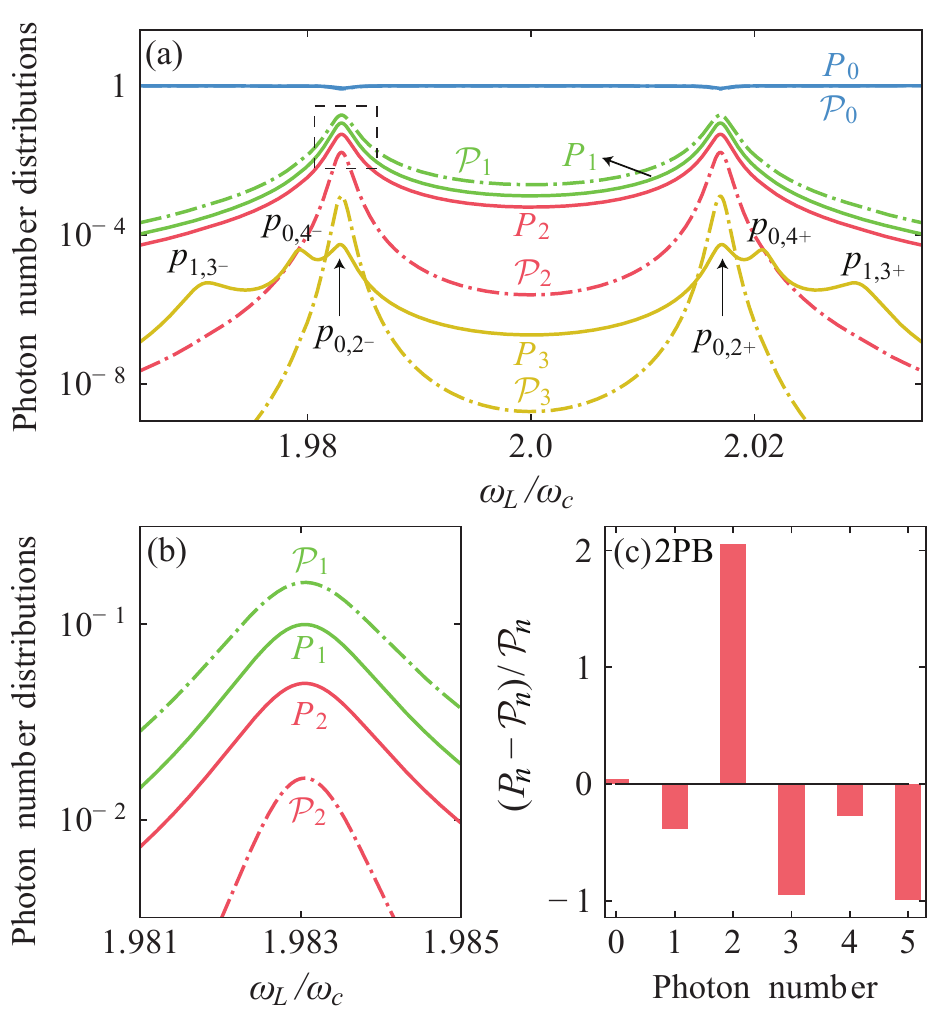}
\caption{(Color online) (a) Photon-number distributions $P_{n=0,1,2,3}$ (solid colored curves) and Poisson distributions $\mathcal{P}_{n=0,1,2,3}$ (dash-dotted colored curves) versus the atomic driving frequency $\omega_{L}/\omega_{c}$ in the resonant case $\omega_{0}/\omega_{c}=2$. (b) Zoomed-in plot of $P_{n=1,2}$ and $\mathcal{P}_{n=1,2}$ versus $\omega_{L}/\omega_{c}$. (c) Relative deviations of the photon-number distribution from the standard Poisson distribution with the same mean photon number located at $\omega_{L}=2\omega_{c}\pm\sqrt{2}J$. Other parameters used are the same as given in Fig.~\ref{Fig6}.}
\label{Fig7}
\end{figure}
The 2PB effect can also be confirmed by comparing the photon-number distributions and the Poisson distributions. In Fig.~\ref{Fig7}(a) we plot $P_{n=0,1,2,3}$ (solid colored curves) and $\mathcal{P}_{n=0,1,2,3}$ (dash-dotted colored curves) as functions of $\omega_{L}/\omega_{c}$ in the resonant case $\omega_{0}/\omega_{c}=2$. Figure~\ref{Fig7}(b) is a zoomed-in plot of $P_{n=1,2}$ and $\mathcal{P}_{n=1,2}$ versus $\omega_{L}/\omega_{c}$. We see in Fig.~\ref{Fig7}(a) that there are two peaks $p_{0,2\pm}$ in the curve of $P_{1}$ (solid green curve) located at $\omega_{L}=2\omega_{c}\pm\sqrt{2}J$, which correspond to the population of the $\vert\varepsilon_{1}\rangle$ induced through the Raman processes $\vert\varepsilon_{0}\rangle\stackrel{\Omega_{L}}{\longrightarrow}\vert\varepsilon_{2\pm}\rangle\stackrel{\kappa}{\longrightarrow}\vert\varepsilon_{1}\rangle$. The physical processes involve the transitions $\vert\varepsilon_{0}\rangle\rightarrow\vert\varepsilon_{2\pm}\rangle$ at the atomic driving frequency $\omega_{L}=2\omega_{c}\pm\sqrt{2}J$, and the decay process $\vert\varepsilon_{2\pm}\rangle\rightarrow\vert\varepsilon_{1}\rangle$. In the curve of $P_{2}$ (solid red curve), we see that there are two peaks $p_{0,2\pm}$ located at $\omega_{L}=2\omega_{c}\pm\sqrt{2}J$, i.e., the two-photon resonance transitions. In the curve of $P_{3}$ (solid yellow curve), we see that there are six peaks ($p_{0,2\pm}$, $p_{0,4\pm}$, and $p_{1,3\pm}$) located at $\omega_{L}=2\omega_{c}\pm\sqrt{2}J$, $2\omega_{c}\pm\sqrt{3}J$, and $2\omega_{c}\pm\sqrt{6}J$, respectively. The two peaks $p_{0,4\pm}$ are induced by the processes $\vert\varepsilon_{0}\rangle\stackrel{\Omega_{L}}{\longrightarrow}\vert\varepsilon_{2\pm}\rangle\stackrel{\Omega_{L}}{\longrightarrow}\vert\varepsilon_{4\pm}\rangle\stackrel{\kappa}{\longrightarrow}\vert\varepsilon_{3\pm}\rangle$, and the two peaks $p_{1,3\pm}$ are induced by the processes $\vert\varepsilon_{0}\rangle\stackrel{\Omega_{L}}{\longrightarrow}\vert\varepsilon_{2\pm}\rangle\stackrel{\kappa}{\longrightarrow}\vert\varepsilon_{1}\rangle\stackrel{\Omega_{L}}{\longrightarrow}\vert\varepsilon_{3\pm}\rangle$. At the locations of the two-photon resonance transitions, we see that the single- and three-photon probabilities are suppressed because $P_{1}<\mathcal{P}_{1}$ and $P_{3}<\mathcal{P}_{3}$, while the two-photon probability is enhanced because $P_{2}>\mathcal{P}_{2}$. This indicates that the 2PB effect can be observed by driving the atom. To further illustrate the 2PB effect, in Fig.~\ref{Fig7}(c) we display the relative deviations of the photon-number distribution from the standard Poisson distribution with the same mean photon number at $\omega_{L}=2\omega_{c}\pm\sqrt{2}J$. We observe that only the value of the two-photon relative population is greater than $0$, i.e., $P_{2}>\mathcal{P}_{2}$, which implies that the 2PB effect can appear in the atom-driving case.

We also study the correlation functions $g^{(2)}(0)$ and $g^{(3)}(0)$ as functions of $J$ ($\kappa$) in the two-photon resonance case. It can be seen that $g^{(2)}(0)$ and $g^{(3)}(0)$ decrease (increase) monotonically with an increase in $J$ ($\kappa$), which means that the 2PB effect is more obvious for a higher coupling strength and the cavity-field decay rate weakens the 2PB effect.

\section{Discussions \label{Discussions}}

Since both two-photon cavity driving and atom driving can increase excitations two by two, it is an interesting question to compare the difference in these two driving cases.
To this end, below we present some discussion of the photon statistics when the cavity field is driven by a two-photon physical process. In this case, the driving Hamiltonian of the system is
\begin{equation}
\hat{H}_{d}^{\prime\prime}=\Omega_{l}(\hat{a}^{\dagger2}e^{-i\omega_{l}t}+\hat{a}^{2}e^{i\omega_{l}t}),
\end{equation}
where $\Omega_{l}$ and $\omega_{l}$ are the driving strength and driving frequency, respectively. In this case, the total Hamiltonian of the system reads $\hat{H}_{\text{sys}}^{\prime\prime}=\hat{H}_{\text{2pJC}}+\hat{H}_{d}^{\prime\prime}$. In a rotating frame defined by the unitary operator $\exp[-i\omega_{l}t(\hat{a}^{\dagger}\hat{a}+\hat{\sigma}_{z})/2]$, the Hamiltonian of the system becomes
\begin{eqnarray}
\hat{H}_{\text{sys}}^{\prime\prime(I)}&=&\Delta^{\prime\prime}_{c}\hat{a}^{\dagger}\hat{a}+\Delta^{\prime\prime}_{0}\hat{\sigma}_{+}\hat{\sigma}_{-}+J(\hat{a}^{\dagger2}\hat{\sigma}_{-}+\hat{\sigma}_{+}\hat{a}^{2})\nonumber\\
&&+\Omega_{l}(\hat{a}^{\dagger2}+\hat{a}^{2}),
\end{eqnarray}
where $\Delta^{\prime\prime}_{c}=\omega_{c}-\omega_{l}/2$ ($\Delta^{\prime\prime}_{0}=\omega_{0}-\omega_{l}$) is the detuning of the cavity-field (atomic) frequency with respect to the driving frequency.

\begin{figure}
\center
\includegraphics[width=0.47 \textwidth]{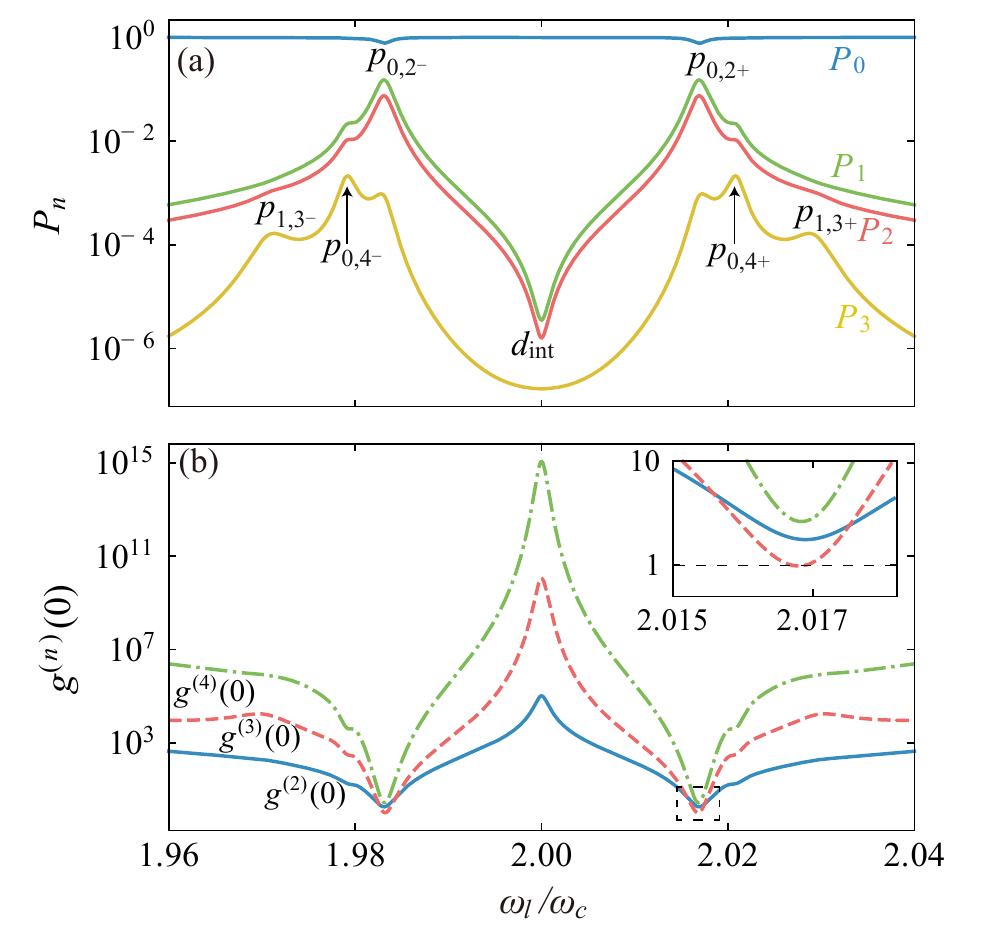}
\caption{(Color online) (a) Photon-number distributions $P_{n=0,1,2,3}$ versus cavity-field driving frequency $\omega_{l}/\omega_{c}$ in the resonant case $\omega_{0}/\omega_{c}=2$. (b) Correlation functions $g^{(2)}(0)$ (solid blue curve), $g^{(3)}(0)$ (dashed red curve), and $g^{(4)}(0)$ (dash-dotted green curve) versus the cavity-field driving frequency $\omega_{l}/\omega_{c}$. Other parameters used are $J/\omega_{c}=0.012$, $\kappa/\omega_{c}=\gamma/\omega_{c}=0.001$, and $\Omega_{l}/\kappa=0.4$.}
\label{Fig8}
\end{figure}

To compare the photon statistics of the two-photon JC model in both the two-photon cavity-driving and the atom-driving cases. In Fig.~\ref{Fig8}(a), we plot the photon-number distributions $P_{n=0,1,2,3}$ as functions of the cavity-field driving frequency $\omega_{l}/\omega_{c}$ in the resonant case $\omega_{0}=2\omega_{c}$ when the cavity field is driven by the two-photon physical process. We find that the locations of these resonance peaks in $P_{n=1,2,3}$ are the same as those in the atom-driving case [see Fig.~\ref{Fig7}(a)]. However, the difference is that there is a dip $d_{\text{int}}$ located at $\omega_{l}/\omega_{c}=2$ in the curve of $P_{2}$ ($P_{1}$), which is induced by destructive quantum interference between the two transition paths. To observe the 2PB effect more clearly, the correlation functions $g^{(2)}(0)$ (solid blue curve), $g^{(3)}(0)$ (dashed red curve), and $g^{(4)}(0)$ (dash-dotted green curve) are plotted versus the cavity-field driving frequency $\omega_{l}/\omega_{c}$ in Fig.~\ref{Fig8}(b). It is found that the correlation functions $g^{(n=2,4)}(0)>1$ and $g^{(3)}(0)\approx1$ at $\omega_{l}=2\omega_{c}\pm\sqrt{2}J$, which means that the 2PB effect cannot occur in the two-photon cavity-driving case. However, the 2PB effect can be observed in the atom-driving case [see Fig.~\ref{Fig6}(e)]. In addition, we find that the PIT effect induced by quantum interference occurs at $\omega_{l}=2\omega_{c}$, i.e., $g^{(n)}(0)>1$ for $n=2,3,4$.

\section{Conclusion \label{conclusion}}

In conclusion, we have studied the multiphoton blockade and PIT effects of the two-photon JC model in both the cavity-field-driving and the atom-driving cases. We have obtained the analytical results of the correlation functions by perturbatively solving the equations of motion for these probability amplitudes. These analytical results are confirmed by numerically solving the quantum master equation including both the cavity-field and the atomic dissipations in the truncated Hilbert space. We have found that the 1PB and PIT effects can be observed in this system for single-photon cavity-field driving, while the 2PB effect cannot occur. In particular, we have shown that the 1PB effect can be enhanced by the destructive quantum interference effect between the two paths in the off-resonant case. Furthermore, we have found that the 2PB effect can be induced by driving the atom, while the 1PB effect cannot be observed because the single-photon transition is forbidden in this case. We have also discussed the two-photon cavity-field-driving case and found that the 2PB effect cannot occur. Our results will pave the way for the study of multiphoton quantum correlation and multiphoton quantum coherent devices.

\begin{acknowledgments}
J.-Q.L. would like to thank Prof. Chi-Kwong Law, Prof. Yu-xi Liu, Prof. Qing-Hu Chen, and Dr. S. Felicetti for helpful discussions. J.-Q.L. is supported in part by the National Natural Science Foundation of China (Grants No.~11822501, No.~11774087, and No.~11935006) and Hunan Science and Technology Plan Project (Grant No.~2017XK2018). J.-F.H. is supported in part by the National Natural Science Foundation of China (Grants No.~12075083 and No.~11505055), Scientific Research Fund of Hunan Provincial Education Department (Grant No.~18A007), and Natural Science Foundation of Hunan Province, China (Grant No.~2020JJ5345).
\end{acknowledgments}

\appendix*

\section{Derivation of the two-photon JC model in a superconducting circuit}

\begin{figure}
\center
\includegraphics[width=0.35 \textwidth]{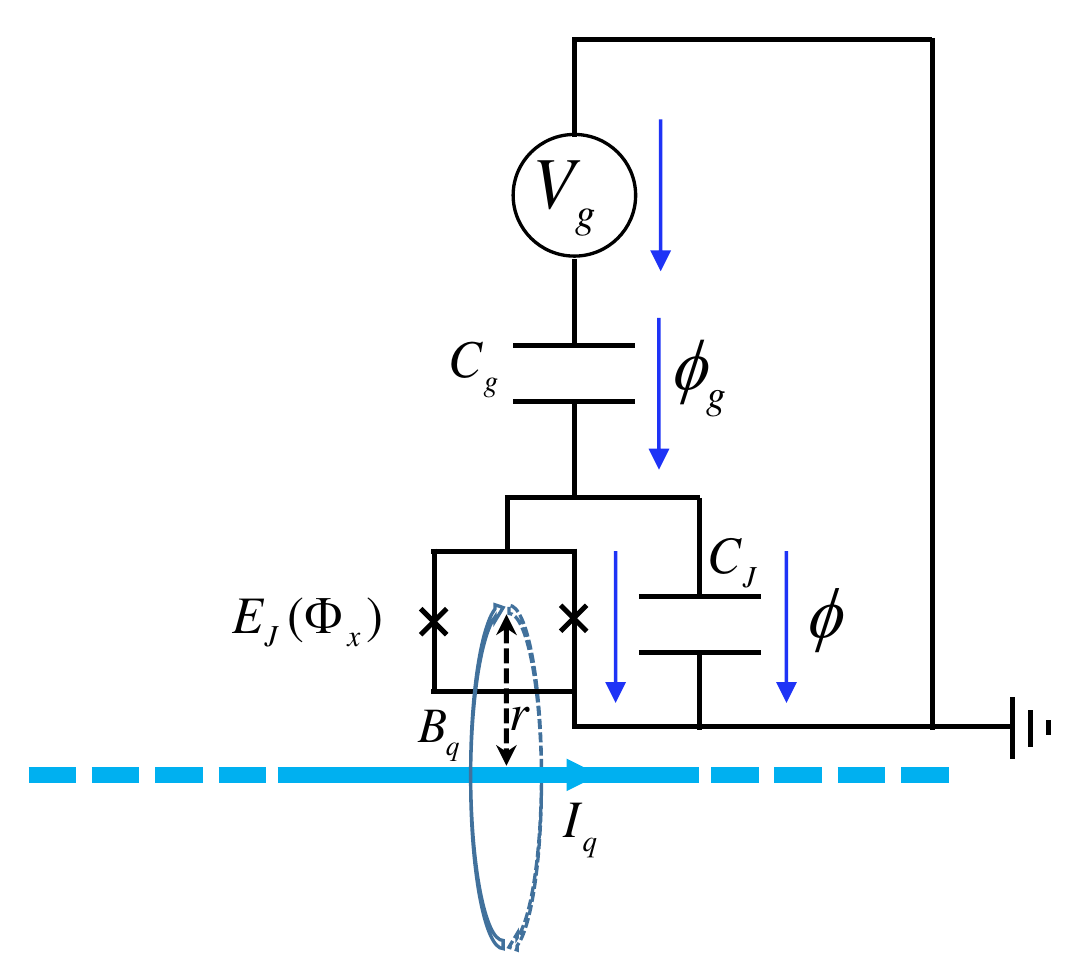}
\caption{Schematic of the superconducting quantum circuit: a superconducting charge qubit (a split-Cooper-pair box) coupled to a superconducting transmission-line resonator.}
\label{Fig9}
\end{figure}

In this Appendix, we present the detailed derivation of the implementation of the two-photon JC model with a superconducting quantum circuit, which is composed of a superconducting charge qubit (a split Cooper-pair box) coupled to a superconducting transmission-line resonator (see Fig.~{\ref{Fig9}}). In this circuit, a superconducting quantum interference device (SQUID; a superconducting loop including two identical Josephson junctions) is connected to a gate voltage $V_{g}$ through a gate capacitance $C_{g}$. Here the SQUID can be described by a single equivalent Josephson junction with a tunable Josephson energy $E_{J}(\Phi_{x})$ and the capacitance $C_{J}$, where $\Phi_{x}$ is the total magnetic flux threading through the superconducting loop. In this circuit, the electrostatic energy plays the role of the kinetic energy, which is given by
\begin{equation}
T=\frac{1}{2}C_{g}\dot{\Phi}_{g}^{2}+\frac{1}{2}C_{J}\dot{\Phi}^{2},
\end{equation}
where $\Phi_{g}$ and $\Phi$ are the generalized magnetic fluxes associated with the phase drops $\phi_{g}$ and $\phi$ across the gate capacitance $C_{g}$ and the SQUID. The relations between the generalized magnetic fluxes and the phase drops are defined by $\phi_{g}=2\pi\Phi_{g}/\Phi_{0}$ and $\phi=2\pi\Phi/\Phi_{0}=(\phi_{1}+\phi_{2})/2$, where $\Phi_{0}$ is the magnetic flux quantum, and $\phi_{1}$ and $\phi_{2}$ are the phases over these two Josephson junctions. The relation between the phase $\phi_{g}$ related to the biased flux $\Phi_{g}$ and the two phases $\phi_{1}$ and $\phi_{2}$ of the two junctions is given by $\phi_{g}=\phi_{1}-\phi_{2}$. The Josephson energy is identified as the potential energy of this system. In this circuit, the Josephson energy of this split Cooper-pair box reads
\begin{equation}
U=-2E_{J}^{0}\cos\left(\frac{\pi\Phi_{x}}{\Phi_{0}}\right)\cos\left( \frac{2\pi}{\Phi_{0}}\Phi\right),
\end{equation}
where $E_{J}^{0}$ is the Josephson energy of a single junction.

Based on the circuit, we have the relation $V_{g}+\dot{\Phi}_{g}+\dot{\Phi}=0$; then the Lagrangian of the system can be written as
\begin{eqnarray}
L&=&T-U=\frac{1}{2}\left(C_{g}+C_{J}\right)\dot{\Phi}^{2}+C_{g}V_{g}\dot{\Phi}\nonumber\\
&&+2E_{J}^{0}\cos\left(\frac{\pi\Phi_{x}}{\Phi_{0}}\right)\cos\left(\frac{2\pi}{\Phi_{0}}\Phi\right)+\frac{1}{2}C_{g}V_{g}^{2}.
\end{eqnarray}
We introduce the momentum canonically conjugate to $\Phi$ as
\begin{equation}
P=\frac{\partial L}{\partial\dot{\Phi}}=\left(C_{g}+C_{J}\right)\dot{\Phi}+C_{g}V_{g},
\end{equation}
then the Hamiltonian of the system can be obtained by the Legendre transformation as~\cite{nakahara2008Quantum}
\begin{align}
H_{\textrm{qub}}&=P\dot{\Phi}-L\notag\\
&=\frac{E_{C}}{2}\left(n-n_{g}\right)^{2}-2E_{J}^{0}\cos\left(\frac{\pi\Phi_{x}}{\Phi_{0}}\right)\cos\phi-\frac{1}{2}C_{g}V_{g}^{2},\nonumber\\
\end{align}
where we introduce the Cooper-pair number $n=P/2e$, the gate-voltage-induced Cooper-pair number $n_{g}=C_{g}V_{g}/(2e)$, the charging energy of one $2$e Cooper pair $E_{C}=4e^{2}/C_{\Sigma}$ with $C_{\Sigma}=C_{g}+C_{J}$ being the total gate capacitance, and the phase difference $\phi=2\pi\Phi/\Phi_{0}$ associated with the SQUID.

The quantization of this superconducting circuit can be performed by introducing the commutative relation between the number operator $\hat{n}$ and the phase operator $\hat{\phi}$ as $[\hat{\phi},\hat{n}]=i$.
Then we can express the Hamiltonian in the eigenrepresentation of the number operator $\hat{n}$ as
\begin{eqnarray}
\hat{H}_{\textrm{qub}}&=&\frac{E_{C}}{2}\sum_{n\in Z}\left(n-n_{g}\right)^{2}\left\vert n\right\rangle\left\langle n\right\vert-E_{J}^{0}\cos\left(\frac{\pi\Phi_{x}}{\Phi_{0}}\right)\nonumber\\
&&\times\sum_{n\in Z}\left(\left\vert n+1\right\rangle\left\langle n\right\vert+\left\vert n\right\rangle\left\langle n+1\right\vert \right)-\frac{1}{2}C_{g}V_{g}^{2},\qquad
\end{eqnarray}
where we have used the relations $\hat{n}=\sum_{n\in Z}n\left\vert n\right\rangle\left\langle n\right\vert$ and $\cos\hat{\phi}=\frac{1}{2}\sum_{n\in Z}\left(\left\vert n+1\right\rangle\left\langle n\right\vert +\left\vert n\right\rangle \left\langle n+1\right\vert\right)$ ($Z$ denoting the integer set). In this work, we consider the case where this circuit works in the charge qubit regime $E_{C}\gg E_{J}$.
In particular, we choose the gate charge in the vicinity of $1/2$, so that the states $\vert 0\rangle$ and $\vert 1\rangle$ have almost-degenerate energies. In this case, other states have higher energies and can be
ignored in our discussion. Then the Hamiltonian becomes
\begin{eqnarray}
\hat{H}_{\textrm{qub}}=\frac{E_{C}}{2}(n_{g}-1/2)\hat{\sigma}_{z}-E_{J}^{0}\cos\left(\frac{\pi\Phi_{x}}{\Phi_{0}}\right)\hat{\sigma}_{x},\label{Hapauliqubit}
\end{eqnarray}
where we introduce the Pauli operators $\hat{\sigma}_{z}=\vert 0\rangle\langle 0\vert-\vert 1\rangle\langle 1\vert$ and $\hat{\sigma}_{x}=\vert 0\rangle\langle 1\vert+\vert 1\rangle\langle 0\vert$. In addition, we have neglected the constant term $E_{C}(1-2n_{g}+2n_{g}^{2})/4-C_{g}V_{g}^{2}/2$ in Eq.~(\ref{Hapauliqubit}).

To realize a two-photon JC model including the superconducting charge qubit driving, we consider the case where the magnetic flux threading the loop of the SQUID consists of three parts: (i) the magnetic flux $\Phi_{e}$ is created by a classical current, (ii) the magnetic flux $\Phi_{q}$ is created by the current in the quantized transmission-line resonator, and (iii) the magnetic flux $\Phi_{s}(t)$ is created by the driving field of the superconducting charge qubit. Then the magnetic flux threading the loop can be expressed as
\begin{equation}
\Phi_{x}=\Phi_{e}+\Phi_{q}+\Phi_{s}(t).
\end{equation}
Here $\Phi_{q}=BS$, with $S$ being the area of the superconducting loop. Since the dimension of the SQUID is much smaller than the length of the transmission-line resonator, we can treat the resonator as a line with an infinite length, then the magnetic field created by the current in the transmission-line resonator can be expressed as $B=\mu_{0} I_{q}/(2\pi r)$, where $\mu_{0}$ is the permeability of free space and $r$ is the distance between the transmission line and the loop. Note that it is reasonable to assume that the magnetic field threading the loop is identical because the dimension of the SQUID loop is much smaller than the distance $r$. The magnetic flux created by the driving field is $\Phi_{s}(t)=\Phi_{s0}\cos(\omega_{s}t)$, where $\omega_{s}$ is the driving frequency.

We assume that the length of the transmission-line resonator is $l$, and the capacitance and inductance per unit length are $C_{0}$ and $L_{0}$. When the free spectrum range of this resonator is large enough, we can focus on one electromagnetic field mode (with resonance frequency $\omega_{c}$) in this resonator. Then the current associated with this mode can be written as
\begin{equation}
I_{q}(x,t)=\sqrt{\frac{\hbar\omega_{c}}{lL_{0}}}(\hat{a}+\hat{a}^{\dagger})\sin\left(\frac{n\pi x}{l}\right),\hspace{0.5 cm}x\in[0,l],
\end{equation}
where $n$ is the index characterizing the considered mode. We assume that the charge qubit is placed at the peak of the field-mode function such that $\sin(n\pi x_{q}/l)=1$, where $x_{q}$ is the location of this charge qubit. Then the flux modulation becomes
\begin{equation}
\cos\left(\frac{\pi\Phi_{x}}{\Phi_{0}}\right)=\cos\left[\frac{\pi\Phi_{e}}{\Phi_{0}}+\phi_{q}(\hat{a}+\hat{a}^{\dagger})+\phi_{s}\cos(\omega_{s}t)\right],\qquad
\end{equation}
where we have introduced the parameter
\begin{equation}
\phi_{q}=\pi\frac{1}{\Phi_{0}}\frac{\mu_{0}}{2\pi r}S\sqrt{\frac{\hbar\omega_{c}}{lL_{0}}},\quad \phi_{s}=\pi\frac{1}{\Phi_{0}}\Phi_{s0},
\end{equation}
which are assumed to be small parameters. We choose a proper classical magnetic flux $\Phi_{e}$ such that $\cos(\pi\Phi_{e}/\Phi_{0})=1$; then we have the approximate relation by expanding the cosine function up to the second order of $\phi_{q}$:
\begin{eqnarray}
\cos\left(\frac{\pi\Phi_{x}}{\Phi_{0}}\right)&=&\cos[\phi_{q}(\hat{a}+\hat{a}^{\dagger})+\phi_{s}\cos(\omega_{s}t)]\nonumber\\
&\approx&1-\frac{1}{2}[\phi_{q}(\hat{a}+\hat{a}^{\dagger})+\phi_{s}\cos(\omega_{s}t)]^{2}.\quad
\end{eqnarray}
Including the free Hamiltonian and the driving Hamiltonian (with the driving strength $\Omega$ and driving frequency $\omega_{d}$), the total Hamiltonian becomes
\begin{eqnarray}
\hat{H}&=&\omega_{c}\hat{a}^{\dagger}\hat{a}+\Omega(\hat{a}^{\dagger}e^{-i\omega_{d}t}+\hat{a}e^{i\omega_{d}t})+\frac{\omega_{0}}{2}\hat{\sigma}_{z}\nonumber\\
&&+J(\hat{a}+\hat{a}^{\dagger})^{2}\hat{\sigma}_{x}+\Omega_{L}(e^{2i\omega_{s}t}+e^{-2i\omega_{s}t})\hat{\sigma}_{x}\nonumber\\
&&-J_{x}\hat{\sigma}_{x}+J_{c}(\hat{a}^{\dagger}+\hat{a})(e^{i\omega_{s}t}+e^{-i\omega_{s}t})\hat{\sigma}_{x},
\end{eqnarray}
where we introduce the parameters $\omega_{0}=E_{C}(n_{g}-1/2)$, $\Omega_{L}=E_{J}^{0}\phi_{s}^{2}/8$, $J=E_{J}^{0}\phi_{q}^{2}/2$, $J_{x}=E_{J}^{0}(1-\phi_{s}^{2}/4)$, and $J_{c}=E_{J}^{0}\phi_{q}\phi_{s}/2$.

We consider the system working under the parameter condition
\begin{eqnarray}
\omega_{0}&\gg& J-J_{x},\hspace{0.5 cm} \omega_{0}\gg 2J\vert n_{a}\vert,\hspace{0.5 cm} \omega_{0}+2\omega_{c}\gg J,\nonumber\\
\Omega_{L}&\ll&\omega_{0}+2\omega_{s},\hspace{0.4 cm} J_{x}\ll\omega_{0}-\omega_{c}-\omega_{s},
\end{eqnarray}
where $n_{a}$ is the maximal photon number involved in mode $a$; then by the rotating-wave approximation, we have the approximate Hamiltonian
\begin{eqnarray}
\hat{H}_{\text{app}}&=&\omega_{c}\hat{a}^{\dagger}\hat{a}+\frac{\omega_{0}}{2}\hat{\sigma}_{z}+J(\hat{\sigma}_{+}\hat{a}^{2}+\hat{a}^{\dagger 2}\hat{\sigma}_{-})\nonumber\\
&&+(\Omega \hat{a}^{\dagger}e^{-i\omega_{d}t}+\Omega_{L}\hat{\sigma}_{+}e^{-i\omega_{L}t}+\text{H.c.}),
\end{eqnarray}
with $\omega_{L}=2\omega_{s}$. This Hamiltonian describes the two-photon JC model including both the electromagnetic-field-driving and the qubit-driving terms. Here the two drivings can be controlled separately by choosing the proper fields.

%

\end{document}